\documentclass[epjST]{svjour}
\usepackage{graphicx} %you should use graphicx instead of graphics, if you want to use graphic option
\usepackage{hyperref}
\usepackage{amsmath} %\spilt
\usepackage{gensymb} %\degree
\usepackage[numbers,sort&compress]{natbib}

\newcommand{\sqrtnn} {\sqrt{s_{\mathrm{NN}}}}
\newcommand{\lr}[1]{\left\langle #1\right\rangle}

\begin{document}
\title{Probe nuclear structure using the anisotropic flow at the Large Hadron Collider}
\author{Zhiyong Lu\inst{1} \and Mingrui Zhao\inst{1}\inst{2} \and Jiangyong Jia\inst{3}\inst{4} \and You Zhou\inst{2}\fnmsep\thanks{\email{you.zhou@cern.ch}}}
\institute{China Institute of Atomic Energy, China \and Niels Bohr Institute, University of Copenhagen, Denmark \and Department of Chemistry, Stony Brook University, USA \and Physics Department, Brookhaven National Laboratory, Upton, NY 11976, USA}
\abstract{
Recent studies have shown that the shape and radial profile of the colliding nuclei have strong influences on the initial condition of the heavy ion collisions and the subsequent development of the anisotropic flow. Using A Multi-Phase Transport model (AMPT) model, we investigated the impact of nuclear quadrupole deformation $\beta_2$ and nuclear diffuseness $a_0$ of $^{129}$Xe 
on various of flow observables in Xe--Xe collisions at $\sqrtnn =$ 5.44 TeV. We found that $\beta_2$ has a strong influence on central collisions while $a_0$ mostly influences the mid-central collisions. The relative change of flow observables induced by a change in $\beta_2$ and $a_0$ are also found to be insensitive to the values of parameters controlling the strength of the interaction among final state particles. Our study demonstrates the potential for constraining the initial condition of heavy ion collisions using future system scans at the LHC.}

%In ultra-relativistic heavy-ion collisions, the distinctive anisotropic behaviors exhibited by the final-state-produced particles serve as a promising avenue for probing the underlying nuclear structure. In this study, we incorporate the nuclear deformation and nuclear diffuseness attributes of $^{129}$Xe into A Multi-Phase Transport model (AMPT), to explore how different nuclear structures affect the results in Xe--Xe collisions at $\sqrtnn =$ 5.44 TeV. We systematically investigate many flow observables, including the flow coefficients $v_n$, nonlinear flow modes $v_{n,mk}$, $\rho_{n,mk}$, $\chi_{n,mk}$, and the normalized symmetric cumulants NSC$(m,n)$. Throughout our analysis, the parameters $\beta_2$ and $a_0$, which describe the nuclear deformation and diffuseness, emerge as pivotal influencers, affecting the flow observables in different centrality ranges. The insights unveiled by the presented AMPT studies reveal the potential avenues for future nuclear structure studies in the heavy-ion experiments at the LHC.} %end of abstract
%
\maketitle
\section{Introduction}
\label{intro}
Ultra-relativistic heavy-ion collisions conducted at both the Relativistic Heavy-Ion Collider (RHIC) and the Large Hadron Collider (LHC) provide a pivotal platform for the comprehensive study of Quantum Chromodynamics (QCD) across both perturbative and nonperturbative regimes. These collisions afford the remarkable opportunity to recreate a novel state of matter, quark-gluon plasma (QGP), characterized by extreme temperatures and densities in the early stages of high-energy heavy-ion interactions. Over the past two decades, a dedicated endeavor has been directed towards extracting precise insights into the properties and the dynamic evolution of QGP~\cite{Shuryak:1980tp,Shuryak:1978ij,Lacey:2006bc,Molnar:2008xj,Muller:2012zq,Drescher:2007cd,Heinz:2013th,Molnar:2001ux,Song:2017wtw,Teaney:2003kp,Xu:2007jv}.
Anisotropic flow, which quantifies the anisotropic expansion of the produced particles, has been a powerful tool for QGP studies~\cite{Muller:2012zq,Drescher:2007cd,Heinz:2013th,Molnar:2001ux,Song:2017wtw,Teaney:2003kp,Xu:2007jv}. It is characterized by the Fourier coefficients $v_n$ of the azimuthal particle distribution~\cite{Voloshin:1994mz}:
\begin{equation}
    f(\varphi) = \frac{1}{2\pi}\left[1+2\sum\limits_{n=1}\limits^{\infty}{v_n\cos[n(\varphi-\Psi_n)]}\right]
    \label{FourierSeries}
\end{equation}
where $\varphi$ is the azimuthal angle of the final particles, $\Psi_n$ is the $n_{th}$-order flow symmetry plane, and $v_n$ is called flow coefficient.
From Eq(\ref{FourierSeries}), the flow coefficient $v_n$ can be defined as:
\begin{equation}
    v_n = \left\langle\cos[n(\varphi-\Psi_n)]\right\rangle.
    \label{FlowCoefficient}
\end{equation}
Here, the angular bracket $\langle\space\rangle$ denotes the average over all particles in one event.
The $v_n$ and flow angle $\Psi_n$ are the magnitude (amplitude) and angle (orientation) of the flow vector, defined as: 
\begin{equation}
V_n \equiv v_n \mathrm{e}^{i n\Psi_n}
\end{equation} 
Systematic measurements on  $v_n$, event-by-event fluctuations of $v_n$, and correlations of different flow coefficients $v_n$, $v_m$, $v_k$ have been previously reported in Refs.~\cite{Niemi:2012aj,Bilandzic:2013kga,ATLAS:2015qwl,Qian:2016pau,Zhu:2016puf}. By performing the Bayesian fit on the extensive flow data, critical information on the temperature dependence of shear and bulk viscosity over entropy density ratios of the QGP, $\eta/s(T)$ and $\zeta/s(T)$, can be extracted~\cite{Bernhard:2019bmu,JETSCAPE:2020mzn,Nijs:2020ors,Parkkila:2021tqq}. 

In addition, the flow measurements give direct access to the event-averaged initial-state shape of the nuclear overlap region and their event-by-event fluctuations. For typical heavy-ion collisions, the nuclear density profile in the initial-state can be described by Woods-Saxon distribution:
\begin{equation}
\begin{split}
    \rho(r,\theta,\phi) &= \frac{\rho_0}{1+e^{[r-R(\theta,\phi)]/a_0}},\\
    R(\theta,\phi) &= R_0(1+\beta_2[\cos\gamma Y_{2,0}+\sin\gamma Y_{2,2}]+\beta_3\sum_{m=-3}^3\alpha_{3,m}Y_{3,m}+\beta_4\sum_{m=-4}^4\alpha_{4,m}Y_{4,m})
    \label{WoodsSaxon}
\end{split}
\end{equation}
where $a_0$ denotes the nuclear diffuseness, while $R_0$ represents the half-width radius. The nuclear surface, denoted as $R(\theta,\phi)$, is expanded in terms of spherical harmonics $Y_{n,m}$, where we retain terms up to $n=4$ as expressed in the Eq.~(\ref{WoodsSaxon}). Furthermore, $\beta_2$, $\beta_3$, and $\beta_4$ stand for the quadrupole, octupole, and hexadecapole deformation parameters, respectively. The parameter $\gamma$ characterizes the triaxial shape, depicting any imbalance present within the axes of the spheroid. Analogous to $\gamma$, $\alpha_{3,m}$ and $\alpha_{4,m}$ describe the inequality of axes and satisfy the normalization condition.

In recent years, various observables have been investigated for their sensitivities to nuclear structure parameters in heavy-ion collisions~\cite{Jia:2021tzt,Zhang:2021kxj,Giacalone:2021udy,Jia:2022qgl,Magdy:2022cvt,Jia:2022qrq,Xu:2021uar,Jia:2021qyu}. As mentioned above, the anisotropic flow reflects the initial spatial anisotropies in the overlap region of the colliding nucleus. Thus, it serves as an ideal probe of initial conditions and can be utilized for the nuclear structure study. The flow coefficient $v_n$ has been found to be sensitive to the deformation, characterized by deformation parameters $\beta_n$, in $^{96}$Ru--$^{96}$Ru, $^{96}$Zr--$^{96}$Zr,$^{238}$U--$^{238}$U, $^{197}$Au--$^{197}$Au collisions~\cite{Jia:2021tzt,Zhang:2021kxj,Giacalone:2021udy,Jia:2022qgl,Magdy:2022cvt}. Beyond $v_n$, the spotlight extends to multi-particle cumulants of $v_n$ and nonlinear flow, underlining their potential for discerning the parameters $\beta_n$, $a_0$, and $R_0$~\cite{Magdy:2022cvt, Jia:2022qrq, Jia:2022qgl}. Moreover, the mean transverse momentum of the produced charged hadrons, denoted as $[p_T]$, which reflects the initial overlap region's size, is a valuable probe for exploring neutron skin thickness and nuclear deformation~\cite{Xu:2021uar}. The fluctuations of $v_n$ and $[p_T]$, as well as the correlations between them, quantified by Pearson correlation coefficient(PCC) and denoted as $\rho(v_n^2, [p_T])$, emerge as a nuanced avenue for constraining deformation parameters $\beta_n$ and the triaxial parameter $\gamma$~\cite{Jia:2021qyu}.

Almost all these studies are conducted within the RHIC energies (GeV energy scale), while a similar study at the LHC energies ranges (TeV energy scale) is still lacking at the moment. In particular, $^{129}$Xe is the nucleus believed to have quadrupole and triaxial deformation~\cite{ALICE:2018yvr}, determined from low-energy nuclear theory and experiments~\cite{Tsukada:2017llu, Fischer:1974aaa, Kumar:1972zza, Poves:2019byh, Cline:1986ik, Morrison:2020azy}. Nevertheless, only very few selected flow observables, such as $v_{n}\{2\}$ and $\rho(v_n^2, [p_T])$, have been studied so far~\cite{ALICE:2018lao,ALICE:2021gxt}. The realm of systematic investigations targeting more intricate flow observables, based on multi-particle correlations in the final state that potentially tie into the many-body interactions within nuclei prior to collisions, are currently unavailable at the LHC. 

This paper will present comprehensive investigations of the nuclear structure using various flow observables, such as flow coefficients, flow fluctuations, correlations between flow coefficients, and nonlinear flow modes, based on the AMPT model simulations. We also study how the final state effects influence these flow observables to ensure that the nuclear parameters can be constrained without being biased by the final state effects.

\section{A Multi-Phase Transport Model}
\label{sec:1}
A Multi-Phase Transport (AMPT) Model~\cite{Lin:2004en} holds widespread application in the realm of ultra-relativistic nuclear collisions for investigating the initial conditions and transport characteristics of Quark-Gluon Plasma (QGP)~\cite{Bhaduri:2010wi,Guo:2019joy,Haque:2019vgi,Lin:2004en,Ma:2016fve,Magdy:2020bhd,Nasim:2010hw,Xu:2010du,Xu:2011fe}. This paper employs the AMPT model incorporating the string melting scenario. The model encompasses a sequence of processes, including the initial conditions of parton production, interactions among partons, hadronization through coalescence, and, finally, hadronic rescattering. More specifically, the nucleons are generated using the HIJING model~\cite{Wang:2000bf} to establish their spatial and momentum distributions, after which they convert into partons. The interactions among these partons are governed by Zhang's Parton Cascade model (ZPC)~\cite{Zhang:1997ej}. In this model, the parton-scattering cross-section $\sigma$ can be characterized by the following:
\begin{equation}
    \sigma = \frac{9\pi\alpha_s^2}{2\mu^2}
    \label{parton-scattering cross-sections}
\end{equation}
where $\alpha_s$ is the QCD coupling constant and $\mu$ is the screening mass. This cross-section delineates the dynamic expansion of the QGP phase. Subsequent to the parton cascade, partons combine to form hadrons, termed hadronization, employing a coalescence model~\cite{Chen:2005mr}. Following hadronization, the interactions among the resulting hadrons in the final state are elucidated by the ART model~\cite{Li:1995pra}.

The AMPT simulations for the Xe--Xe collisions are employed by parameterizing the nucleon density profile with Woods-Saxon distribution shown in Eq (\ref{WoodsSaxon}). To investigate the effect of nuclear deformation and diffuseness, and also the sensitivity to the system's dynamic evolution, several sets of values for $\beta_2$, $\gamma$, $a_0$, and $\sigma$ are used for comparisons. Here we use sets 1--4 to study the effect of nuclear deformations, i.e., by changing $\beta_2$ from 0 (spherical) to 0.18 (deformed, obtained from Ref.~\cite{ALICE:2018yvr}) and changing $\gamma$ from 0 (prolate) to 27$\degree$ (triaxial) to 60$\degree$ (oblate). Also, comparing the results from sets 2 ($a_0 = 0.57$ from Ref.~\cite{ALICE:2018yvr}) and 5 ($a_0 = 0.492$ from Ref.~\cite{Bally:2021qys}) can provide information for nuclear diffuseness. For the impact of transport properties, we use $\sigma=3.0$ mb (set 6)\cite{Ma:2014pva,Bzdak:2014dia} instead of $\sigma=6.0$ mb (set 3)\cite{Feng:2016emh}. The observables are shown in centrality dependence, where the centrality in this study is determined by the impact parameter in Xe--Xe collisions. More detailed information concerning the input parameters of the AMPT model can be found in table~\ref{tab:Xe--XeAMPT}:
\begin{table}[hbt]
    \caption{parameter sets used in this study}
    \label{tab:Xe--XeAMPT}
    \begin{center}
    \begin{tabular}{ccccc}
    \hline
    set & $\beta_2$ & $\gamma$ & $a_0$ & $\sigma$\\
    \hline
	Set 1 & 0 & 0 & 0.57 & 6.0 mb \\
	Set 2 & 0.18 & 0 & 0.57 & 6.0 mb \\
	Set 3 & 0.18 & 27$\degree$ & 0.57 & 6.0 mb \\
	Set 4 & 0.18 & 60$\degree$ & 0.57 & 6.0 mb \\
	Set 5 & 0.18 & 0 & 0.492 & 6.0 mb \\
	Set 6 & 0.18 & 27$\degree$ & 0.57 & 3.0 mb \\
    \hline
    \end{tabular}
    \end{center}
\end{table}

\section{Analysis details}
\label{sec:2}
\subsection{Observables}
\label{sec:2:1}
Experimentally, flow coefficients cannot be obtained directly via Eq.~(\ref{FlowCoefficient}) but from two- and multi-particle  correlations/cumulants\cite{Bilandzic:2013kga,Bilandzic:2010jr,Borghini:2000sa,Moravcova:2020wnf}:
\begin{equation}
    v_n\lbrace 2\rbrace\equiv \sqrt{c_n\lbrace 2\rbrace}
    \label{Definition TwoParticleCumulants}
\end{equation}
where $c_n\lbrace 2\rbrace$ is the two-particle cumulant.
In this analysis,  $v_{2}\{2\}$,$v_{3}\{2\}$,$v_{4}\{2\}$ are studied. In addition, the four-particle cumulants of $v_n$, denoted as $v_{n}\{4\}$, can be obtained via four-particle cumulants $c_n\{4\}$:
\begin{equation}
    v_n\{4\}\equiv\sqrt[4]{-c_n\{4\}}.
    \label{FourParticleV2}
\end{equation}
The higher order cumulants of $v_n$ are defined in a similar way, noted as $v_n\{6\}$, $v_n\{8\}$, etc.
The two- and multi-particle cumulants of $v_n$ have different contributions from flow fluctuations $\sigma_{v_n}$.
It is well known that for the Gaussian type flow fluctuations and in the case $\sigma_{v_n}\ll\bar{v}_n$, we have~\cite{Voloshin:2007pc}:
\begin{equation}
    \begin{split}
    v_n\{2\}^2&\approx\bar{v}_n^2+\sigma_{v_n}^2,\\
    v_n\{4\}^2&\approx\bar{v}_n^2-\sigma_{v_n}^2,\\
    v_n\{6\}^2&\approx\bar{v}_n^2-\sigma_{v_n}^2,\\
    v_n\{8\}^2&\approx\bar{v}_n^2-\sigma_{v_n}^2,
    \label{FlowFluctuations}
    \end{split}
\end{equation}
here $\sigma_{v_n}$ is the standard deviation of $v_n$ distribution, which represents the event-by-event fluctuations of $v_n$. Then mean flow coefficient $\bar{v}_n$ (which is also known as $v_n$ from the flow symmetry plane) and the flow fluctuation $\sigma_{v_n}$ can be extracted from the combination of $v_n\{2\}$ and $v_n\{4\}$ according to the Eq.~(\ref{FlowFluctuations}):
\begin{equation}
    \begin{split}
    \bar{v}_n &\approx \sqrt{\frac{v_n\{2\}^2+v_n\{4\}^2}{2}},\\
    \sigma_{v_n} &\approx \sqrt{\frac{v_n\{2\}^2-v_n\{4\}^2}{2}}
    \label{Meanv2AndFlowFluctuations}
    \end{split}
\end{equation}

For central and semi-central collisions, the lower order flow coefficients $v_{n}$ (for $n=2,3$) are linearly correlated with the initial eccentricity coefficients $\varepsilon_n$~\cite{Niemi:2012aj,Song:2010mg}. While higher harmonic flow $v_{n}$ (for $n>3$) not only has the linear response to the corresponding initial $\varepsilon_n$ but also has contributions from the lower order $\varepsilon_{2}$ and/or $\varepsilon_{3}$ ~\cite{Bhalerao:2014xra,Bhalerao:2013ina,Yan:2015jma}. The latter is called the nonlinear flow mode. For example, $V_4$ and $V_5$ can be decomposed into the linear and nonlinear components:
\begin{equation}
    \begin{split}
    V_4=V^\mathrm{NL}_4+V^\mathrm{L}_4&\approx\chi_{4,22}(V_2)^2+V^\mathrm{L}_4,\\
    V_5=V^\mathrm{NL}_5+V^\mathrm{L}_5&\approx\chi_{5,32}V_2V_3+V^\mathrm{L}_5.
    \label{LinearAndNonlinearPart}
    \end{split}
\end{equation}
Here $V^\mathrm{NL}_n$ and $V^\mathrm{L}_n$ are the nonlinear and linear (or called leftover) components, respectively. Their magnitudes are denoted as $v_{n,mk}$ and $v_{n}^{L}$. Besides, $\chi_{n,mk}$ is the nonlinear coefficient representing the strength of nonlinear response from lower order eccentricities~\cite{Yan:2015jma}. The correlation between different order flow symmetry planes can be studied by calculating the ratio between $v_{n,mk}$ and $v_n\{2\}$~\cite{ALICE:2017fcd}:
\begin{equation}
    \begin{split}
    \rho_{4,22}&=\frac{v_{4,22}}{v_4\{2\}},\\
    \rho_{5,32}&=\frac{v_{5,32}}{v_5\{2\}}
    \label{Rhonmk}
    \end{split}
\end{equation}
$\rho_{4,22}$ and $\rho_{5,32}$ can be used to study the correlations between $\Psi_2$ and $\Psi_4$, as well as the correlations between three planes of $\Psi_2$, $\Psi_3$ and $\Psi_5$. The study of nonlinear flow modes, i.e., $v_{n,mk}$, $\rho_{n,mk}$, $\chi_{n,mk}$ have been performed before, they could provide further constraints on the initial conditions~\cite{Bhalerao:2014xra,Zhou:2015eya,Bilandzic:2013kga}.

The correlations between $v_n^2$ and $v_m^2$ can be quantified via normalized symmetric cumulants NSC$(m,n)$, defined as~\cite{Bilandzic:2013kga}:
\begin{equation}
\mathrm{NSC}(m,n) = 
\frac{\langle v_m^2 \, v_n^2 \rangle - \langle v_m^2\rangle \langle v_n^2 \rangle}{\langle v_m^2\rangle \langle v_n^2 \rangle},
\label{NSC}
\end{equation}
where the angular bracket  $\langle\space\rangle$ represents an average over all events. It allows to study if $v_n^2$ and $v_m^2$ are correlated, anti-correlated or uncorrelated, if NSC$(m,n)$ $>0$, $< 0$ and $=0$, respectively. In this paper, NSC$(3,2)$ and NSC$(4,2)$ will be studied with the AMPT model to see if the results could bring extra information into the initial conditions and the structure of $^{129}$Xe.

\subsection{Multi-particle correlation}
\label{sec:2:2}
All the flow observables introduced in section~\ref{sec:2:1} can be obtained via the multi-particle correlation method \cite{Bilandzic:2013kga,Bilandzic:2010jr,Borghini:2000sa,Moravcova:2020wnf}. To begin with, the flow coefficient $v_n$ can be calculated using two-particle correlations:
\begin{equation}
    v_n\{2\}\equiv \sqrt{c_n\{2\}}=\langle\langle\cos n(\varphi_1-\varphi_2)\rangle\rangle^{1/2},
    \label{TwoParticleCorrelation}
\end{equation}
where $\varphi_1$ and $\varphi_2$ are azimuthal angles from different particles. Double brackets $\langle\langle\space\rangle \rangle$ denote the average over all particles in an event and then the average over all events.

For multi-particle cumulants of $v_n$~\cite{Moravcova:2020wnf}:
\begin{equation}
    \begin{split}
        v_n\{4\} &\equiv \sqrt[4]{-c_n\{4\}},\\
        v_n\{6\} &\equiv \sqrt[6]{\frac{1}{4}c_n\{6\}},\\
        v_n\{8\} &\equiv \sqrt[8]{-\frac{1}{33}c_n\{8\}}
        \label{MultiPartcilevn}
    \end{split}
\end{equation}
where :
\begin{equation}
    \begin{split}
        c_n\{4\} &= \langle v_n^4\rangle-2\langle v_n^2\rangle^2\\
        c_n\{6\} &= \langle v_n^6\rangle
        - 9\langle v_n^4\rangle\langle v_n^2\rangle
        +12\langle v_n^2\rangle^3,\\
        c_n\{8\} &= \langle v_n^8\rangle-16\langle v_n^6\rangle\langle v_n^2\rangle-18\langle v_n^4\rangle^2+144\langle v_n^4\rangle\langle v_n^2\rangle^2-144\langle v_n^2\rangle^4
        \label{MultiPartcileCumulants}
    \end{split}
\end{equation}

and 
\begin{equation}
    \begin{split}
       % \langle v_n^2\rangle &= \langle\langle\cos(n\varphi_1-n\varphi_2)\rangle\rangle,\\
        \langle v_n^4\rangle &= \langle\langle\cos(n\varphi_1+n\varphi_3-n\varphi_2-n\varphi_4)\rangle\rangle,\\
        \langle v_n^6\rangle &= \langle\langle\cos(n\varphi_1+n\varphi_3+n\varphi_5-n\varphi_2-n\varphi_4-n\varphi_6)\rangle\rangle,\\
        \langle v_n^8\rangle &= \langle\langle\cos(n\varphi_1+n\varphi_3+n\varphi_5+n\varphi_7-n\varphi_2-n\varphi_4-n\varphi_6-n\varphi_8)\rangle\rangle
        \label{MultiPartcileCumulants2}
    \end{split}
\end{equation}
The magnitude of nonlinear flow mode $v_{n,mk}$ could be obtained via the multi-particle correlations~\cite{Yan:2015jma,ALICE:2017fcd}.
For $n=4,5$, which is studied in this paper, we have:
\begin{equation}
    \begin{split}
    v_{4,22}&=\frac{\langle\langle\cos(4\varphi_1-2\varphi_2-2\varphi_3)\rangle\rangle}{\sqrt{\langle\langle\cos(2\varphi_1+2\varphi_2-2\varphi_3-2\varphi_4)\rangle\rangle}},\\
    v_{5,32}&=\frac{\langle\langle\cos(5\varphi_1-3\varphi_2-2\varphi_3)\rangle\rangle}{\sqrt{\langle\langle\cos(3\varphi_1+2\varphi_2-3\varphi_3-2\varphi_4)\rangle\rangle}}
    \label{VnmkInCorrelation}
    \end{split}
\end{equation}
They quantify the magnitude of the nonlinear mode in high-order flow coefficients. By subtracting $v_{4,22}$ and $v_{5,32}$ from $v_4$, and $v_5$, respectively, we can easily calculate the magnitude of linear modes:
\begin{equation}
    \begin{split}
        v_4^\mathrm{L} &= \sqrt{v_4^2\{2\} - v_{4,22}^2},\\
        v_5^\mathrm{L} &= \sqrt{v_5^2\{2\} - v_{5,32}^2}
    \end{split}
\end{equation}
Nonlinear coefficient $\chi_{n,mk}$ also describes the nonlinear contributions but is independent of $v_2$ and $v_3$. It can be derived by taking the ratio of $v_{n,mk}$ and corresponding lower-order $v_n$($n=2,3$):
\begin{equation}
    \begin{split}
    \chi_{4,22}&=\frac{v_{4,22}}{\sqrt{\lr{v_2^4}}}=\frac{\langle\langle\cos(4\varphi_1-2\varphi_2-2\varphi_3)\rangle\rangle}{{\langle\langle\cos(2\varphi_1+2\varphi_2-2\varphi_3-2\varphi_4)\rangle\rangle}},\\
    \chi_{5,32}&=\frac{v_{5,32}}{\sqrt{\lr{v_2^2 v_3^2}}}=\frac{\langle\langle\cos(5\varphi_1-3\varphi_2-2\varphi_3)\rangle\rangle}{{\langle\langle\cos(3\varphi_1+2\varphi_2-3\varphi_3-2\varphi_4)\rangle\rangle}}
    \label{ChinmkInCorrelation}
    \end{split}
\end{equation}
The $\rho_{n,mk}$, correlation between different order flow symmetry planes can be obtained by taking Eqs.~(\ref{TwoParticleCorrelation}) and (\ref{VnmkInCorrelation}):
\begin{equation}
    \begin{split}
    \rho_{4,22}&=\frac{v_{4,22}}{v_4\{2\}}
    =\frac{\langle\langle\cos(4\varphi_1-2\varphi_2-2\varphi_3)\rangle\rangle}{\sqrt{\langle\langle\cos(2\varphi_1+2\varphi_2-2\varphi_3-2\varphi_4)\rangle\rangle\langle\langle\cos(4\varphi_1-4\varphi_2)\rangle\rangle}},\\
    \rho_{5,32}&=\frac{v_{5,32}}{v_5\{2\}}
    =\frac{\langle\langle\cos(5\varphi_1-3\varphi_2-2\varphi_3)\rangle\rangle}{\sqrt{\langle\langle\cos(3\varphi_1+2\varphi_2-3\varphi_3-2\varphi_4)\rangle\rangle\langle\langle\cos(5\varphi_1-5\varphi_2)\rangle\rangle}}
    \label{RhonmkInCorrelation}
    \end{split}
\end{equation}
Expanding Eq.~(\ref{NSC}) with multi-particle correlations, normalized symmetric cumulants NSC$(m,n)$ can be expressed as: 
\begin{displaymath}
    \mathrm{NSC}(m,n) = 
    \frac{
    \langle\langle \cos(m\varphi_1+n\varphi_3-m\varphi_2-n\varphi_4) \rangle\rangle
    -\langle\langle\cos(m\varphi_1-m\varphi_3)\rangle\rangle\langle\langle\cos(n\varphi_2-n\varphi_4)\rangle\rangle
    }{\langle\langle\cos(m\varphi_1-m\varphi_3)\rangle\rangle\langle\langle\cos(n\varphi_2-n\varphi_4)\rangle\rangle}
    \label{NSCInCorrelation}
\end{displaymath}

All these observables are now expressed in terms of 2- and multi-particle correlations and then can be calculated using the {\it Generic~Framework}~\cite{Bilandzic:2013kga,Huo:2017nms} or its latest implementation {\it Generic~Algorithm}~\cite{Moravcova:2020wnf}.

\section{Results}
\label{sec:3}
\subsection{Study on the nuclear deformation}
\label{sec:3:1}

\begin{figure}[!htb]
    \begin{center}
      \includegraphics[width=\textwidth]{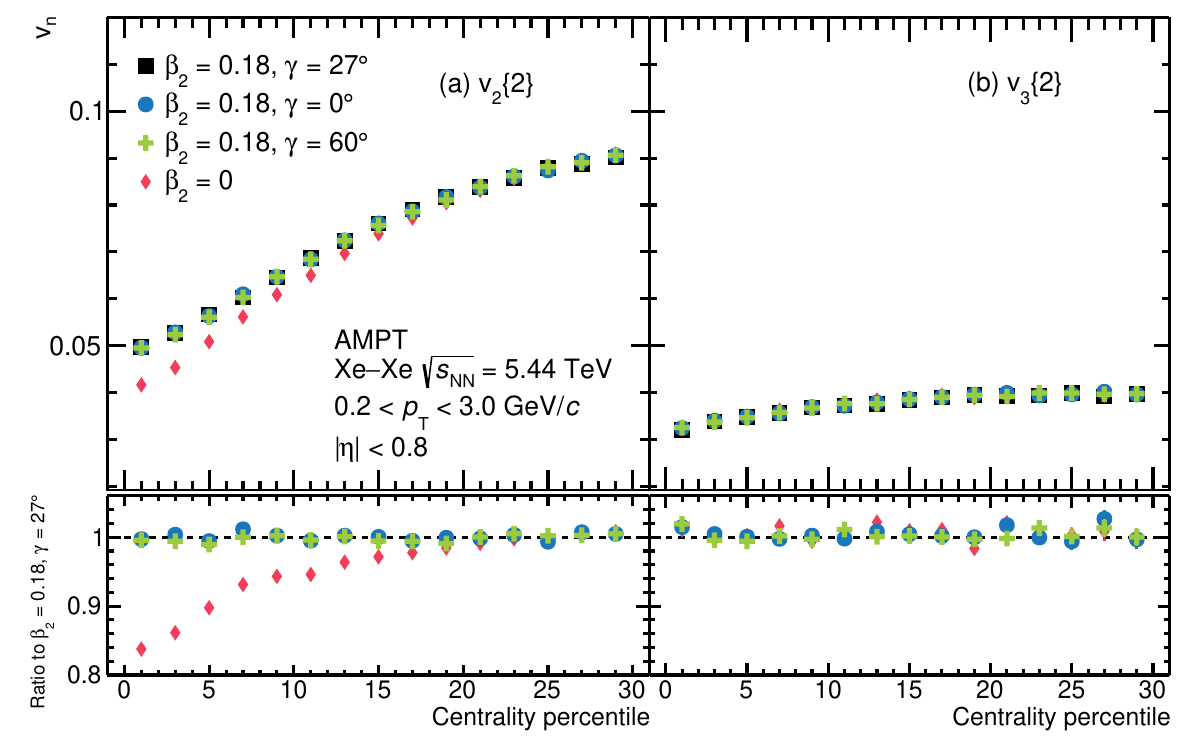}
      \caption[.]{Centrality dependence of $v_n\{2\}(n=2,3)$ in Xe--Xe collisions at $\sqrt{s_\mathrm{NN}}$ = 5.44 TeV in AMPT.}
      \label{FigureSet0}
    \end{center}
  \end{figure}

The centrality dependence of $v_{2}\{2\}$ and $v_{3}\{2\}$ in Xe-Xe collisions at 5.44 TeV are shown in Fig.~\ref{FigureSet0}.
Here, $v_{2}\{2\}$ increases significantly in the Ultra-Central Collisions (UCC) region when changing $\beta_2$ from 0 (red diamonds) to 0.18 (the other markers). In the UCC region, where the two colliding nuclei almost fully overlap, the eccentricity of the overlapping region is determined by the shape or nuclear structure of the colliding nuclei. As it's well known, the $\beta_2$ parameter in the Woods-Saxon distribution characterizes an elliptical shape of the Woods-Saxon nucleon density profile. A non-zero $\beta_2$ of $^{129}$Xe enhances the initial eccentricity $\varepsilon_2$ compared to the one with a spherical shape where $\beta_2=0$ and consequently leads to an increase in $v_2$ because of the linear correlation between $\varepsilon_n$ and $v_n$, i.e., $v_n \propto \varepsilon_n$, for $n=$ 2 or 3~\cite{Niemi:2015qia,Schenke:2020uqq}. In Fig.~\ref{FigureSet0}(b), $v_{3}\{2\}$ has negligible differences when changing the $\beta_2$ values, because the triangularity $\varepsilon_3$ of the overlapping region does not depend on $\beta_2$~\cite{Giacalone:2021udy}.

Figure \ref{FigureSet0} also shows consistent results of $v_{n}\{2\}(n=2,3)$ with variations in the triaxial parameter $\gamma$ across the 0--30\% centrality range. This is not a surprise, as probing the 3-D structure usually requests correlation involving more than two particles. In light of preceding investigations~\cite{Jia:2021tzt}, a potential sensitivity of $\varepsilon_2$ to $\gamma$ was indicated for centrality range 0--0.2\%, as concluded from simulations using the initial-state model. However, the current study refrains from probing such phenomena, relying on the AMPT model's final state information. The intrinsic computational demands inherent to capturing the entirety of the dynamic evolution process within AMPT serve as a significant impediment.

\begin{figure}[!htb]
    \begin{center}
      \includegraphics[width=\textwidth]{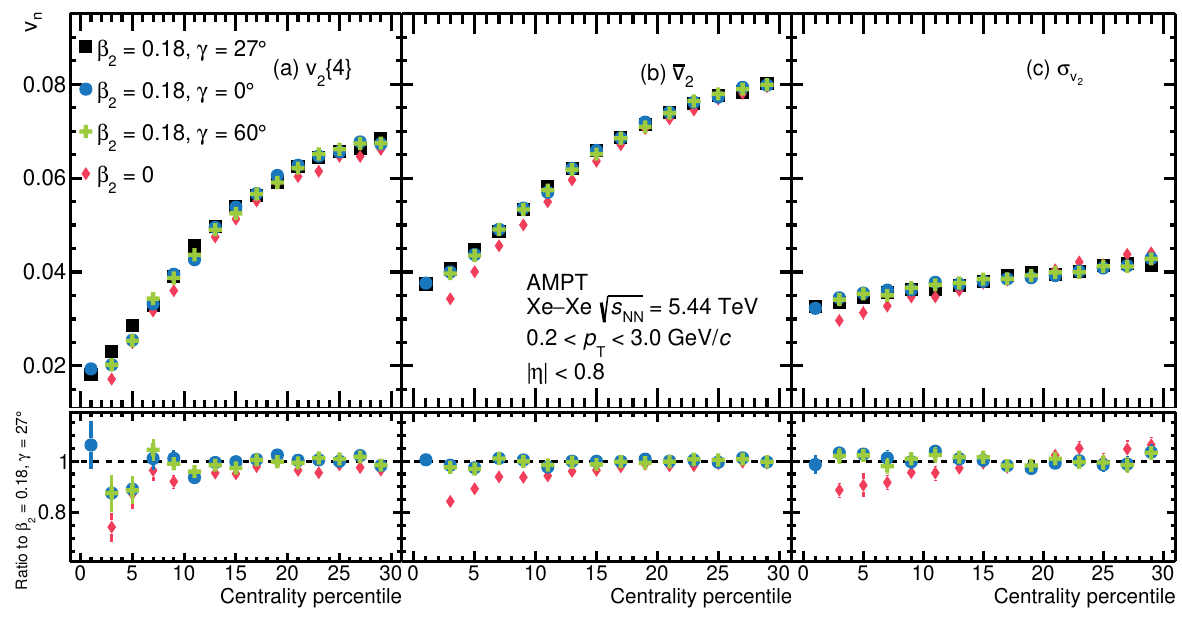}
      \caption[.]{Centrality dependence of $v_2\{4\}, \bar{v}_2, \sigma_{v_2}$ in Xe--Xe collisions at $\sqrt{s_\mathrm{NN}}$ = 5.44 TeV in AMPT.}
      \label{FigureSet1}
    \end{center}
  \end{figure}

The value of $v_{2}\{2\}$ receives the contributions not only from the initial eccentricity $\varepsilon_2$ but also from the event-by-event eccentricity fluctuations. To undertake a more comprehensive exploration of the ramifications imposed by the initial event-by-event eccentricity fluctuations and correspondingly the final state elliptic flow fluctuations, the utilization of both $v_{2}\{2\}$ and $v_{2}\{4\}$ is employed. This combined approach is particularly insightful as these two observables carry opposite contributions from flow fluctuations, as shown in Eqs.~(\ref{FlowFluctuations}). As a result, the centrality dependence of $v_{2}\{4\}$, as well as the $\bar{v}_2$ and $\sigma_{v_2}$ in Xe--Xe collisions at 5.44 TeV are presented in Fig.~\ref{FigureSet1}. As shown in Fig.~\ref{FigureSet1}(a), $v_{2}\{4\}$ exhibits a slight increase when changing $\beta_2$ from 0 (red diamonds) to 0.18 (the other markers), despite the sizable statistical uncertainties in the presented centrality region. It was previously suggested that $v_{2}\{4\}$ is influenced by the nuclear diffuseness $a_0$, with a weak sensitivity to $\beta_2$ using different nuclei at the RHIC energy~\cite{Jia:2022qgl}. This aligns with our AMPT results shown in Figs.~\ref{FigureSet1}(a) and~\ref{FigureA0Set1}(a). In Fig.~\ref{FigureSet1}(b)(c), $\bar{v}_2$ and $\sigma_{v_2}$ have significant increases in UCC region when changing $\beta_2$ from 0 to 0.18. The sensitivity of $\bar{v}_2$ to $\beta_2$ primarily arises from its linear correlation with $\varepsilon_2$. In the case of flow fluctuations $\sigma_{v_2}$, a deformed nucleus assumes varied orientations compared to a spherical shape. These random orientations result in stronger fluctuations of the elliptic flow.
 
In Fig.~\ref{FigureSet1}, $v_{2}\{4\}$, $\bar{v}_2$, and $\sigma_{v_2}$ stay unchanged with variations in $\gamma$.
These observables do not lend themselves to the task of constraining the unique triaxial parameter characterizing the nucleus $^{129}$Xe.

\begin{figure}[!htb]
    \begin{center}
      \includegraphics[width=\textwidth]{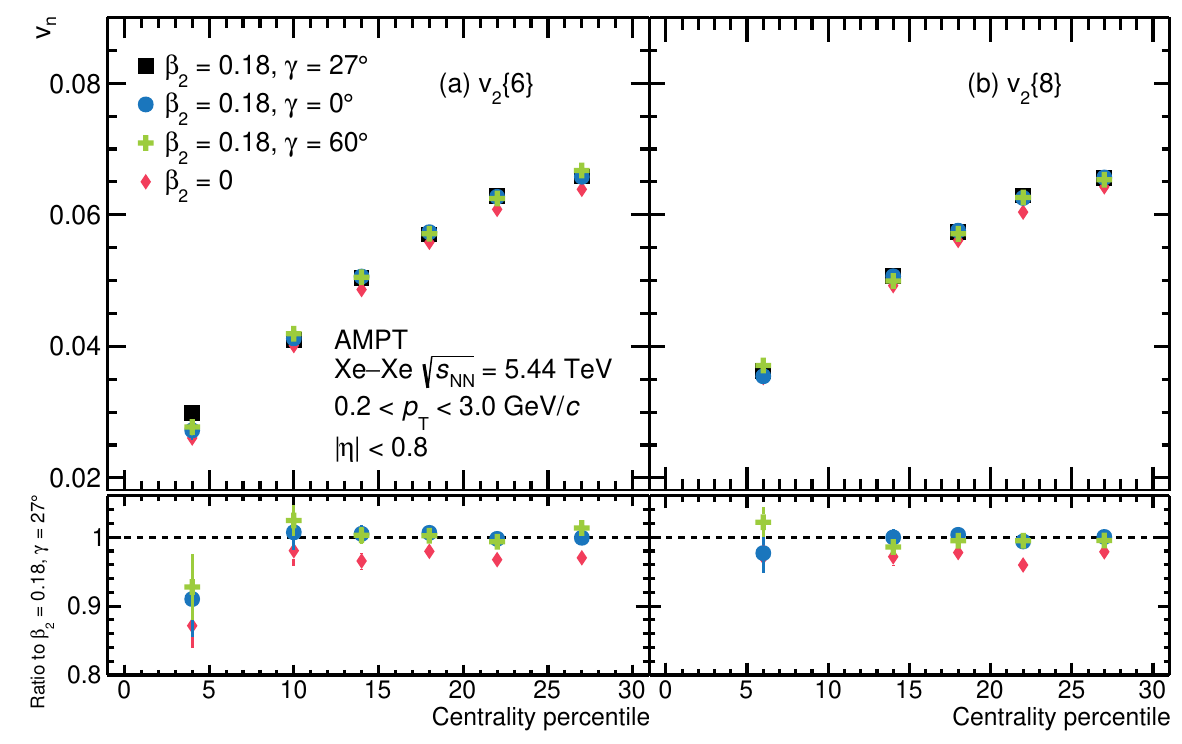}
      \caption[.]{Centrality dependence of $v_2\{6\},v_2\{8\}$ in Xe--Xe collisions at $\sqrt{s_\mathrm{NN}}$ = 5.44 TeV in AMPT.}
      \label{FigureSetv26v28}
    \end{center}
  \end{figure}

An evident sensitivity of $v_{2}\{6\}$ to $\beta_3$ is reported in a previous study~\cite{Jia:2022qgl}. The same work also discusses the correlation between the sensitivity of $\beta_n$ and the number of particles used in the multi-particle correlations. Following the same concept, it is expected that observables like $v_{2}\{6\}$ and $v_{2}\{8\}$ may yield distinctive insights into the nuclear structure, compared to $v_{2}\{2\}$ and $v_{2}\{4\}$ discussed above. The centrality dependence of $v_{2}\{6\}$ and $v_{2}\{8\}$ are shown in Fig.~\ref{FigureSetv26v28}. Amidst the sizable uncertainties, the current study refrains from drawing a firm conclusion regarding the sensitivity of these two observables to either $\beta_2$ or $\gamma$ across the 0--30\% centrality interval.

\begin{figure}[!htb]
    \begin{center}
      \includegraphics[width=\textwidth]{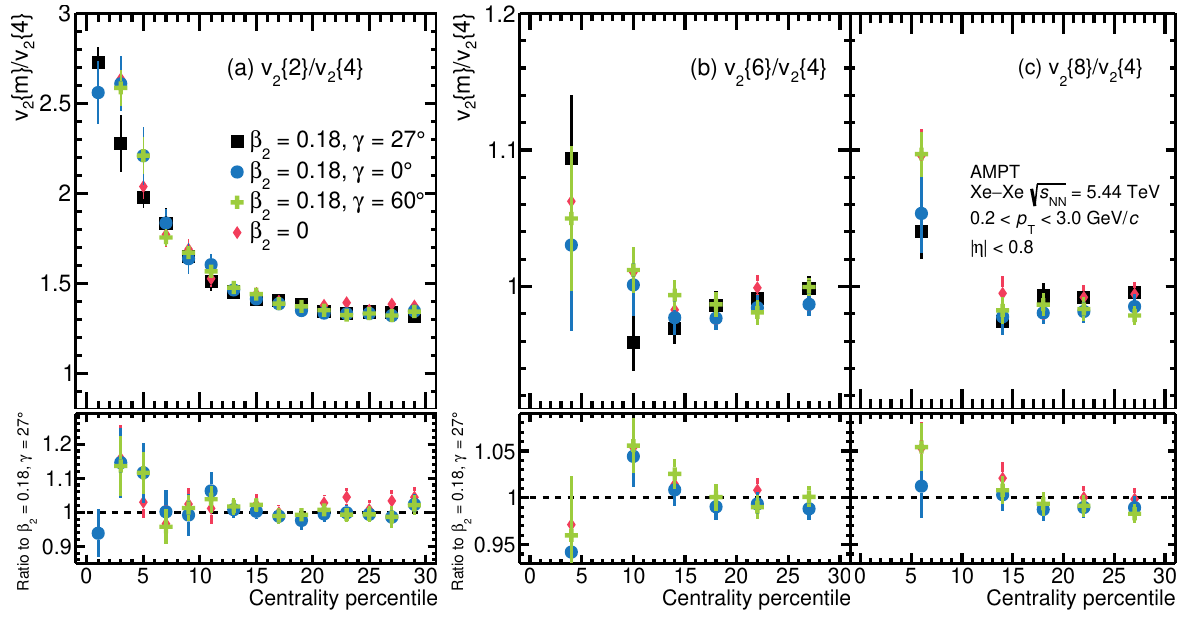}
      \caption[.]{Centrality dependence of $v_2\{2\}/v_2\{4\}$ in Xe--Xe collisions at $\sqrt{s_\mathrm{NN}}$ = 5.44 TeV in AMPT.}
      \label{Figurev22v26v28overv24}
    \end{center}
  \end{figure}

The $v_n$ with multi-particle correlations, i.e., $v_{2}\{4\}$, $v_{2}\{6\}$, and $v_{2}\{8\}$, were reported to exhibit a slight difference at the precision of 1-2\% in Pb--Pb collisions~\cite{ALICE:2018rtz}. This difference is believed to have originated from the deviations from a Bessel-Gaussian shape, particularly a non-zero skewness of the event-by-event $v_2$ distribution. To determine whether a similar pattern persists in Xe--Xe collisions and explore the discrepancies between higher-order multi-particle cumulants, the ratios $v_2\lbrace m\rbrace/v_2\{4\}$ for $m=$ 2, 6, and 8 are presented as a function of centrality in Fig.~\ref{Figurev22v26v28overv24}. The ratio of $v_2\{2\}/v_2\{4\}$ increases dramatically toward the central region in Fig.~\ref{Figurev22v26v28overv24}(a). Such an increase is dominated by flow fluctuations in the most central region. Although the sensitivity of $v_2\{2\}/v_2\{4\}$ to $\beta_2$ and $\gamma$ is covered by uncertainties, a weak sensitivity to $\beta_2$ was observed previously in a similar study in U--U collisions at $\sqrt{s_\mathrm{NN}}$ = 193 GeV~\cite{Magdy:2022cvt}. Subsequent investigations with increased statistics are anticipated to elucidate the presence of this sensitivity. However, the current study is constrained by the substantial computing demands involved. Concerning higher-order cumulants with $m$=6,8, the ratios $v_2\{6\}/v_2\{4\}$ and $v_2\{8\}/v_2\{4\}$ are close to unity within uncertainties and they stay unchanged when varying the nuclear structure configurations $\beta_2$ and $\gamma$.

\begin{figure}[!htb]
    \begin{center}
      \includegraphics[width=\textwidth]{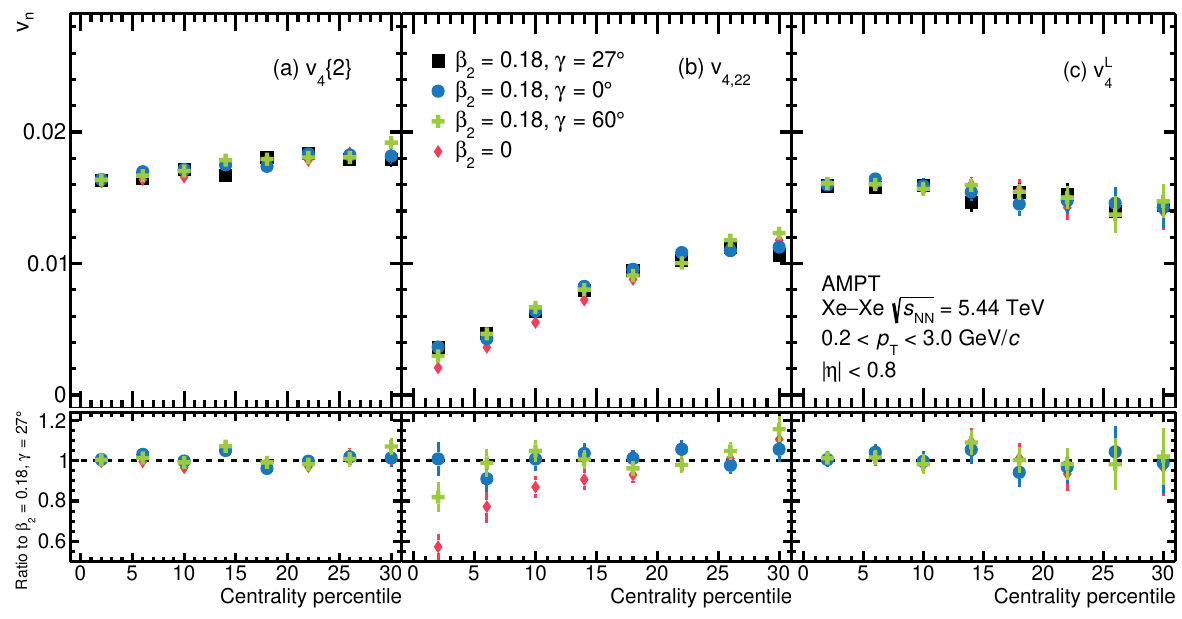}
      \caption[.]{Centrality dependence of nonlinear modes including (a)$v_{4}$, (b)$v_{4,22}$ and (c)Linear $v_{4}$ in Xe--Xe collisions at $\sqrt{s_\mathrm{NN}}$ = 5.44 TeV in AMPT.}
      \label{FigureNonlinearVn4}
    \end{center}
\end{figure}

In the context of the higher harmonic flow ($n>3$), such as $v_4$, it consists of a linear constituent $v^\mathrm{L}_4$ stemming from the initial $\varepsilon_4$ and a nonlinear component $v^\mathrm{NL}_4$ originating from $\varepsilon_2^{2}$. The centrality dependence of $v_4\{2\}$, $v^\mathrm{L}_{4}$, and $v^\mathrm{NL}_4$ or denoted as $v_{4,22}$ are presented in Figure \ref{FigureNonlinearVn4}. In Fig.~\ref{FigureNonlinearVn4}(a)(b), $v_4\{2\}$ and $v_{4,22}$ exhibit an increasing trend moving from the central to the mid-central collision, while $v^\mathrm{L}_{4}$ in Fig.~\ref{FigureNonlinearVn4}(c) has a flat distribution in the presented centrality region. Notably, a comparison of the magnitudes of $v_4\{2\}$ and $v^\mathrm{L}_{4}$ reveals the prevailing dominance of $v^\mathrm{L}_{4}$ over $v_{4,22}$ in central collisions, while in the mid-central region $v_{4,22}$ converges with $v^\mathrm{L}_{4}$. This is consistent with previous measurements in Pb-Pb collisions at the LHC~\cite{ALICE:2017fcd}. The linear component $v_{4}^{L}$ shows an absence of sensitivity to the quadrupole deformation parameter $\beta_2$ across the 0--30\% centrality range. A similar observation has been reported recently in Ref.~\cite{Magdy:2022cvt}, where both $v^\mathrm{L}_4$ and $v_4\{2\}$ show no sensitivity to $\beta_2$. In contrast, the sensitivity of both $v_{4,22}$ and $v^\mathrm{L}_4$ to the hexadecapole deformation parameter $\beta_4$ is seen~\cite{Magdy:2022cvt}, owing to their shared provenance from $\varepsilon_4$. The nonlinear component $v_{4,22}$, as illustrated in Fig.~\ref{FigureNonlinearVn4}(b), manifests a degree of 20\% reductions with zero $\beta_2$ within the UCC region, albeit with a magnitude smaller than that of $v_4\{2\}$. Since the nonlinear flow $v_{n,mk}$ probes a smaller spatial distribution than the flow coefficient $v_n$, it can impose more stringent constraints on initial conditions and nuclear structure parameters. Variations in $\gamma$ parameter show that $v_{4,22}$, $v^\mathrm{L}_4$, and the $v_4\{2\}$ all exhibit diminished sensitivities to $\gamma$ within the 0--30\% centrality range, and thus they are unable to probe the triaxial nuclear structure.

\begin{figure}[!htb]
    \begin{center}
      \includegraphics[width=\textwidth]{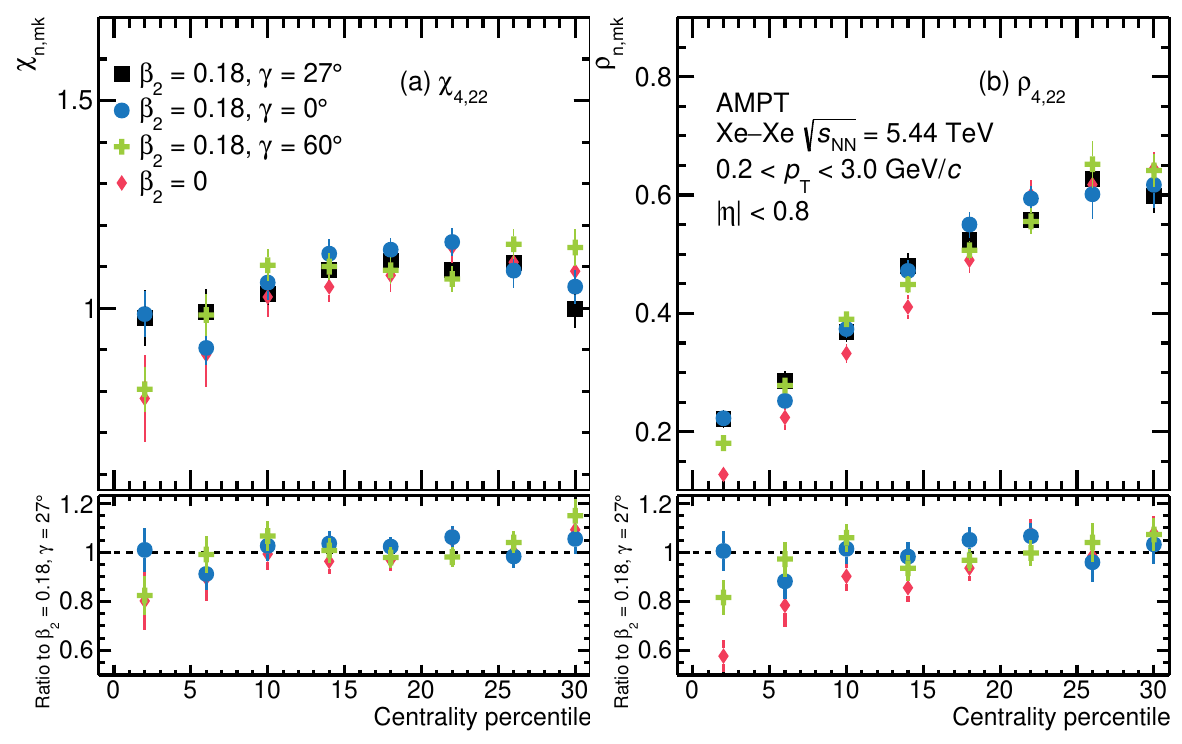}
      \caption[.]{Centrality dependence of nonlinear modes including (a)$\chi_{4,22}$ and (b)$\rho_{4,22}$ in Xe--Xe collisions at $\sqrt{s_\mathrm{NN}}$ = 5.44 TeV in AMPT.}
      \label{FigureNonlinearMode4}
    \end{center}
\end{figure}

The nonlinear coefficient $\chi_{4,22}$, defined in Eq.~(\ref{LinearAndNonlinearPart}), is not solely affected by transport properties but also by initial conditions~\cite{ALICE:2020sup}. It previously exhibits a weak dependence on nuclear deformation $\beta_2$ in central collisions at the RHIC isobar runs~\cite{Jia:2022qrq,Magdy:2022cvt, Zhao:2022uhl}. Verifying whether the same conclusion holds in the context of Xe--Xe collisions remains pertinent. Nonlinear correlation $\rho_{4,22}$, introduced in Eq.~(\ref{Rhonmk}), demonstrates an insensitivity to the influences of final state interactions~\cite{ALICE:2020sup}, rendering it an effective tool for probing initial conditions. In Fig.~\ref{FigureNonlinearMode4}, the centrality dependence of $\chi_{4,22}$ and $\rho_{4,22}$ are presented. In panel (a), $\chi_{4,22}$ stays unchanged with different nuclear structure configurations in the presented centrality region despite the large uncertainties in the central collisions. Considering the relationship $V_4^\mathrm{NL} = \chi_{4,22} \, (V_2)^2$, the sensitivity of $v_{4,22}$ to $\beta_2$ primarily stems from the sensitivity conveyed by $(V_2)^2$. A similar scenario can be observed in the case of $v_{5,32}$ and $\chi_{5,32}$, as shown in Fig.\ref{FigureNonlinearMode5} in the appendix, suggesting that the sensitivity arises from $v_2v_3$. Furthermore, the results in Fig.~\ref{FigureSet0} have already indicated that $v_3$ is insensitive to $\beta_2$, thereby emphasizing that the dominant sensitivity of $v_{5,32}$ to $\beta_2$ emerges from $v_2$. In Fig.~\ref{FigureNonlinearMode4}(b), $\rho_{4,22}$ shows a distinct drop with $\beta_2$=0 in UCC region. This reduction is caused by the $v_{4,22}$ in Fig.~\ref{FigureNonlinearVn4}(b) with $\beta_2$=0, as $\rho_{4,22}=v_{4,22}/v_4\{2\}$ and no such sensitivity is observed in $v_{4}\{2\}$. Turning to a physics view, $\rho_{4,22}$ signifies the correlation between $\Psi_4$ and $\Psi_2$. When the initial geometry is spherical ($\beta_2=0$), $\Psi_4$ and $\Psi_2$ could be arbitrary orientations and thus have negligible correlations. When the geometry has an anisotropy ($\beta_2>0$), they will have specific orientations. As a result, their correlation increases as well. Changing triaxial parameter $\gamma$, the results of $\chi_{4,22}$ and $\rho_{4,22}$ remain consistent within errors, restraining themselves from an ideal probe of the triaxial structure of $^{129}$Xe.

\begin{figure}[!htb]
    \begin{center}
      \includegraphics[width=\textwidth]{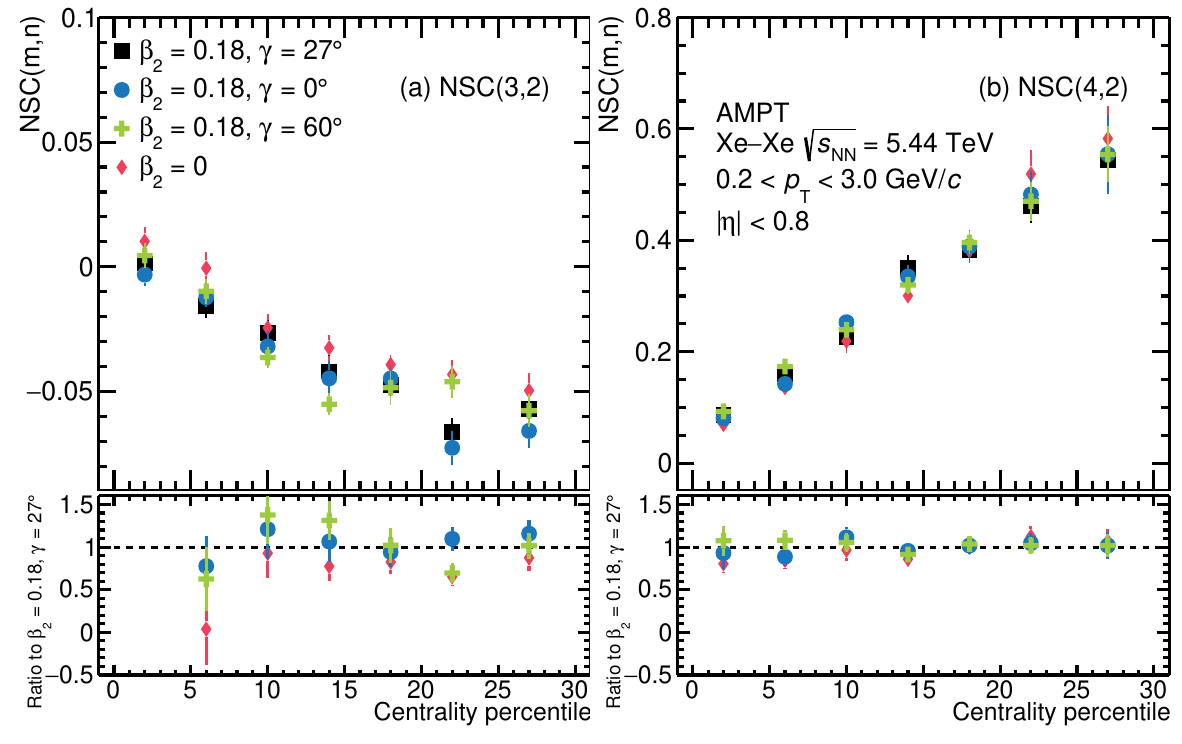}
      \caption[.]{Centrality dependence of NSC$(m,n)$ in Xe--Xe collisions at $\sqrt{s_\mathrm{NN}}$ = 5.44 TeV in AMPT.}
      \label{FigureNSC}
    \end{center}
  \end{figure}

Normalized symmetric cumulant (NSC) quantifies the correlation between different order flow coefficients and has been systematically studied in Pb--Pb collisions~\cite{ALICE:2016kpq,ALICE:2021adw,ALICE:2017kwu}. It was found that NSC$(3,2)$ is insensitive to the dynamic evolution but carries unique sensitivity to initial conditions~\cite{ALICE:2016kpq,ALICE:2021adw}. Thus, it is potentially a good probe of the nuclear structure. In fact, NSC$(3,2)$ has been previously studied in the U--U collisions, showing its sensitivity to $\beta_2$~\cite{Magdy:2022cvt}.
While for NSC$(4,2)$, it is sensitive to both initial conditions and transport properties~\cite{ALICE:2016kpq,ALICE:2021adw}. In Figure \ref{FigureNSC}(a), a tiny (and insignificant) increase can be observed in NSC$(3,2)$ by decreasing the $\beta_2$ value from 0.18 to 0, and no significant variation is observed when changing $\gamma$. This indicates that the anti-correlation between $\langle v_3^2\rangle$ and $\langle v_2^2\rangle$ is reduced by quadrupole deformation and is independent of $\gamma$. For Fig.~\ref{FigureNSC}(b), it is evident that NSC$(4,2)$ exhibits little to no sensitivity to either $\beta_2$ or $\gamma$, suggests that the correlation between $\langle v_4\rangle$ and $\langle v_2\rangle$ remains relatively independent of nuclear deformation.

\subsection{Study on the nuclear diffuseness}
\label{sec:3:2}

The radial profile of the nucleus is determined by either the nuclear diffuseness $a_0$ or the radius $R_0$, both of which impact the initial spatial anisotropy and consequently influence the flow observables. Recent studies have highlighted the significant impact of $a_0$ on various flow observables in the mid-central region (20--60\% centrality)\cite{Jia:2022qgl}, underscoring the importance of verifying this within the context of Xe--Xe collisions using the AMPT model. This section delves into the potential of flow observables in investigating the $a_0$ introduced in Eq.~(\ref{WoodsSaxon}). Generally, a larger nuclear diffuseness $a_0$ results in a significantly increased total hadronic cross-section, leading to a more diffused QGP.

\begin{figure}[!htb]
    \begin{center}
      \includegraphics[width=\textwidth]{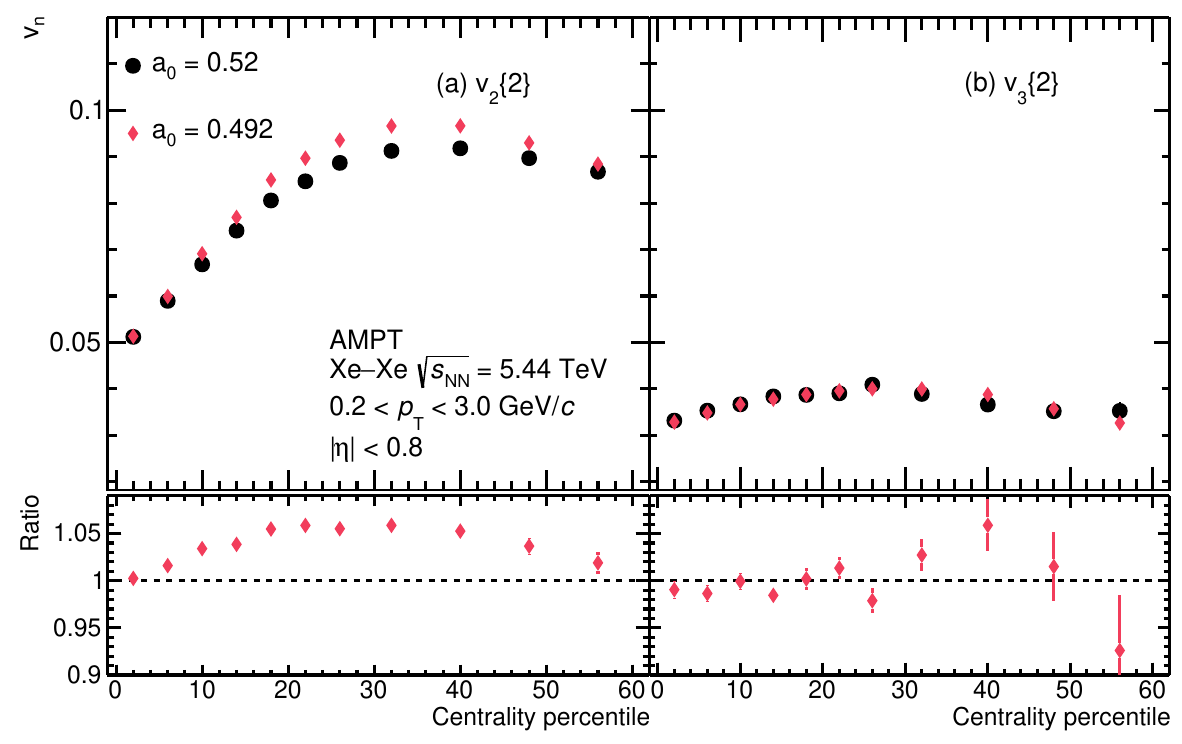}
      \caption[.]{Centrality dependence of $v_n\{2\}(n=2,3)$ in Xe--Xe collisions at $\sqrt{s_\mathrm{NN}}$ = 5.44 TeV in AMPT.}
      \label{FigureA0Setv2v3}
    \end{center}
  \end{figure}

Fig.~\ref{FigureA0Setv2v3} shows the centrality dependence of $v_2\{2\}$ and $v_3\{2\}$ in Xe--Xe collisions at 5.44 TeV using two distinct values of $a_0$. Calculations with larger $a_0$ values (presented by black markers) lead to smaller $v_2$ results in the semi-central collisions, which is agreed with the results in the isobar runs of Ru--Ru/Zr--Zr~\cite{Jia:2022qgl}. Meanwhile, no obvious change in $v_3$ is seen after changing the $a_0$ values across the 0--60\% centrality range.

\begin{figure}[!htb]
    \begin{center}
      \includegraphics[width=\textwidth]{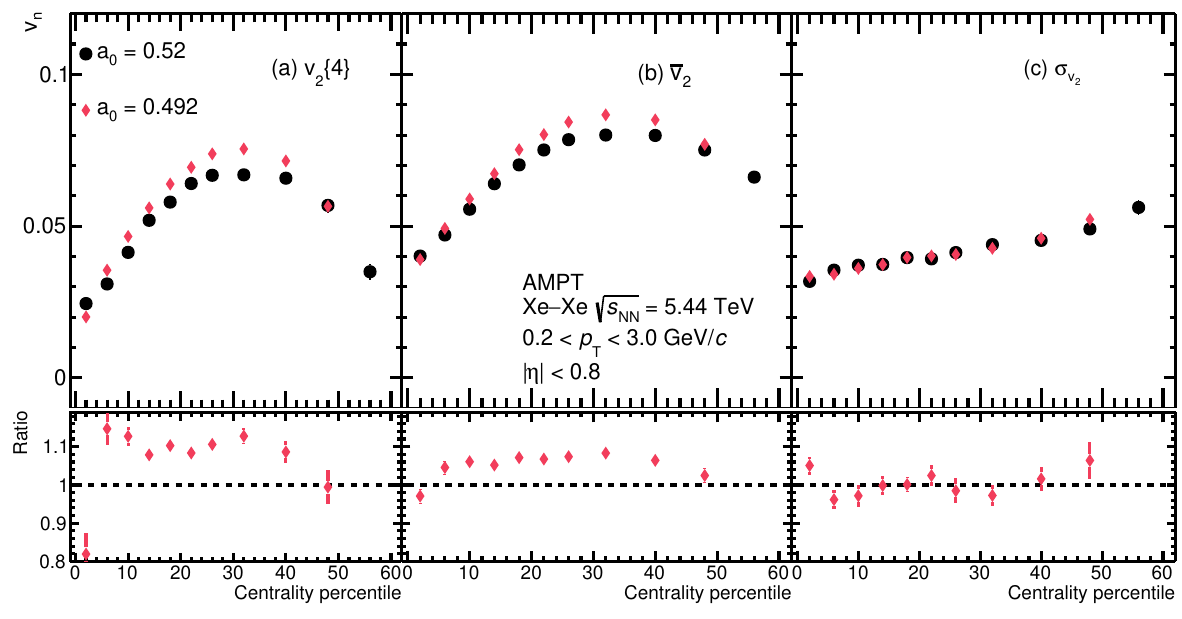}
      \caption[.]{Centrality dependence of $v_2\{4\}, \bar{v}_2, \sigma_{v_2}$ in Xe--Xe collisions at $\sqrt{s_\mathrm{NN}}$ = 5.44 TeV in AMPT.}
      \label{FigureA0Set1}
    \end{center}
  \end{figure}
  
As discussed in section~\ref{sec:3:1}, the increase of $\sigma_{v_2}$ is mainly due to the random orientations of deformed nuclei and is expected to be less influenced by the diffuseness of nuclei. To verify this, the results of $v_{2}\{4\}$, $\bar{v}_2$, and $\sigma_{v_2}$ are presented in Fig.~\ref{FigureA0Set1}. It is observed that $v_{2}\{4\}$ and $\bar{v}_2$ are reduced by larger $a_0$ in the semi-central region, while $\sigma_{v_2}$ remains the same. This observation aligns with our earlier discussion, affirming that the diffuseness primarily impacts the eccentricity rather than its fluctuations.

\begin{figure}[!htb]
    \begin{center}
      \includegraphics[width=\textwidth]{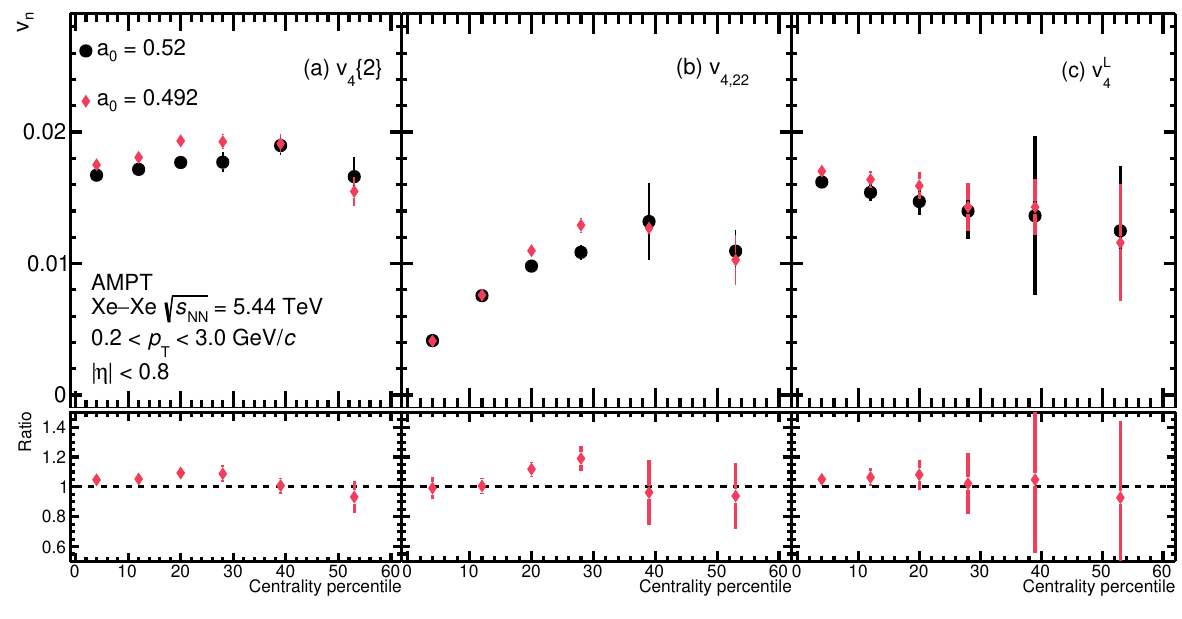}
      \caption[.]{Centrality dependence of nonlinear modes including (a)$v_{4}$, (b)$v_{4,22}$ and (c)Linear $v_{4}$ in Xe--Xe collisions at $\sqrt{s_\mathrm{NN}}$ = 5.44 TeV in AMPT.}
      \label{FigureA0NonlinearVn4}
    \end{center}
\end{figure}

In exploring $a_0$, we can split $v_4$ into its linear and nonlinear components to see how they respond differently to changes in $a_0$. Figure~\ref{FigureA0NonlinearVn4} shows the centrality dependence of $v_4\{2\}$ and its linear and nonlinear components. A larger $a_0$ value reduces both $v_4\{2\}$ and $v_{4,22}$ in 20--40\% centrality. Figure~\ref{FigureA0NonlinearVn4}(c) does not show the sensitivity of $v_4^\mathrm{L}$ to $a_0$ due to the large uncertainties. The reduction of the nonlinear part $v_{4,22}$ affects $v_4\{2\}$. This is different from what we saw for $\beta_2$. Also, it is seen that $v_{4,22}$ is much smaller than $v_4^\mathrm{L}$ in the most central region, so we do not see any effect of $\beta_2$ on $v_4\{2\}$ even though $\beta_2$ increases $v_{4,22}$ significantly in that region. For $a_0$, more sensitivity is found in the mid-central collisions, where the nonlinear component $v_{4,22}$ is about half of the total $v_4\{2\}$, and how it changes with $a_0$ has a bigger effect on $v_4\{2\}$ in this region.

\begin{figure}[!htb]
    \begin{center}
      \includegraphics[width=\textwidth]{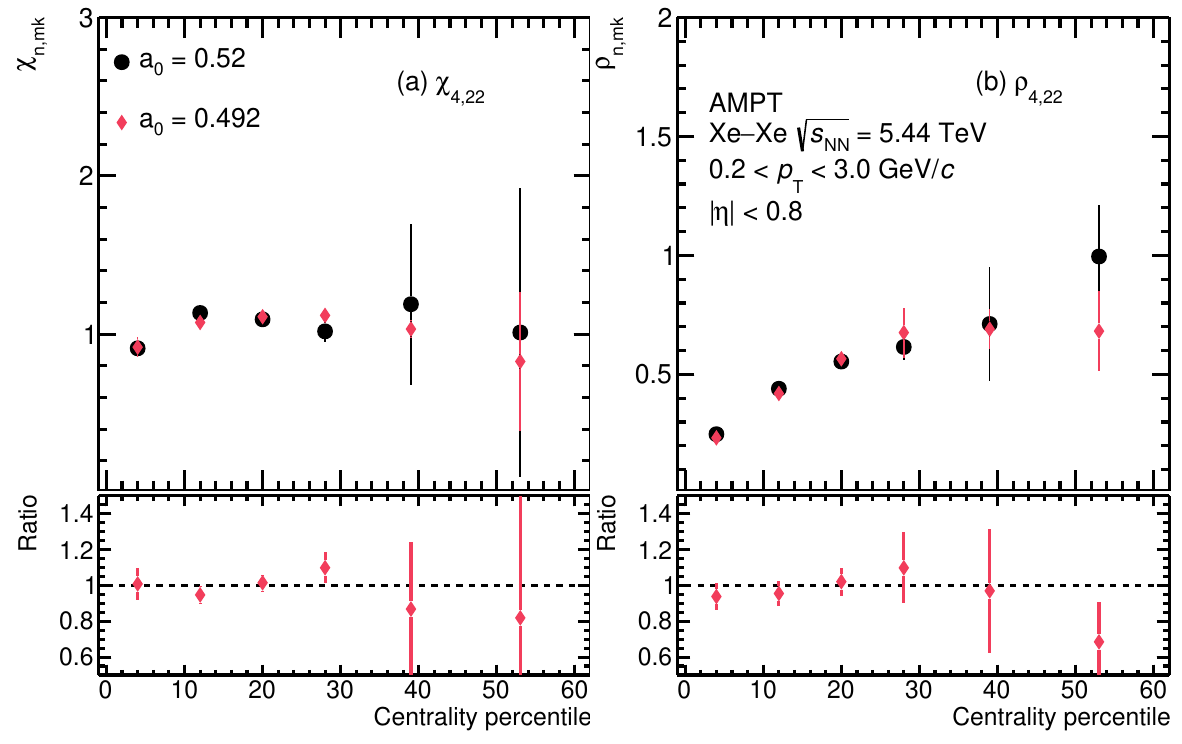}
      \caption[.]{Centrality dependence of nonlinear modes including (a)$\chi_{4,22}$ and (b)$\rho_{4,22}$ in Xe--Xe collisions at $\sqrt{s_\mathrm{NN}}$ = 5.44 TeV in AMPT.}
      \label{FigureA0NonlinearMode4}
    \end{center}
\end{figure}

As also introduced in section~\ref{sec:3:1}, $\chi_{4,22}$ is not only affected by dynamic evolution but also by initial conditions, while $\rho_{4,22}$ only depends on the initial state. It is crucial to check if the nonlinear modes can probe the initial conditions, particularly the diffuseness of nuclei. Here $\chi_{4,22}$ and $\rho_{4,22}$ with different $a_0$ are presented in Fig.~\ref{FigureA0NonlinearMode4}. Neither of them is affected by $a_0$ in 0--60\% centrality region. As $\rho_{4,22}=v_{4,22}/v_4\{2\}$, the sensitivity to $a_0$ is, to a large extent, canceled by taking this ratio. The insensitivity of $\chi_{4,22}$ to $a_0$ can lead to similar conclusion as in the discussion of $\beta_2$ that the sensitivity of $v_{4,22}$ to $a_0$ is mainly due to the $V_2^2$ term in $V_4^\mathrm{NL}\approx\chi_{4,22}(V_2)^2$.

Because of the limited statistics, there are large uncertainties in the results of NSC(3,2) and NSC(4,2), and no firm conclusion about how sensitive they are to the $a_0$ parameter can be drawn. Thus, the results are not presented and discussed here but added in the appendix.

\subsection{Influence from the transport properties}
\label{sec:3:3}

The results shown in this study have indicated that $v_{2}\{2\}$, $\bar{v}_2$, $\sigma_{v_2}$, $v_{4,22}$ and $\rho_{4,22}$ can potentially probe the nuclear deformation, meanwhile $v_{2}\{2\}$, $v_{2}\{4\}$, $\bar{v}_2$, $v_{4}\{2\}$, $v_{4,22}$ have the ability to study nuclear diffuseness $a_0$. In order to achieve unbiased constraints on the nuclear structure parameters in experiments, one must ensure that the chosen observables are only sensitive to the initial conditions and not influenced by the dynamic evolution of the created system. In particular, it has been found in the study at RHIC isobar runs that the ratio observables in Zr--Zr and Ru--Ru can largely cancel out the effects of final state interactions and thus reflect mainly the impact from the initial stages. Here we present the mentioned observables ($v_{2}\{2\}$, $v_{2}\{4\}$, $\bar{v}_2$, $\sigma_{v_2}$,  $v_{4}\{2\}$, $v_{4,22}$, $\rho_{4,22}$) in the ratio of Xe--Xe/Pb--Pb, checking if the ratio of two collision system minimizes the influence of dynamic evolution.

\begin{figure}[!htb]
    \begin{center}
      \includegraphics[width=\textwidth]{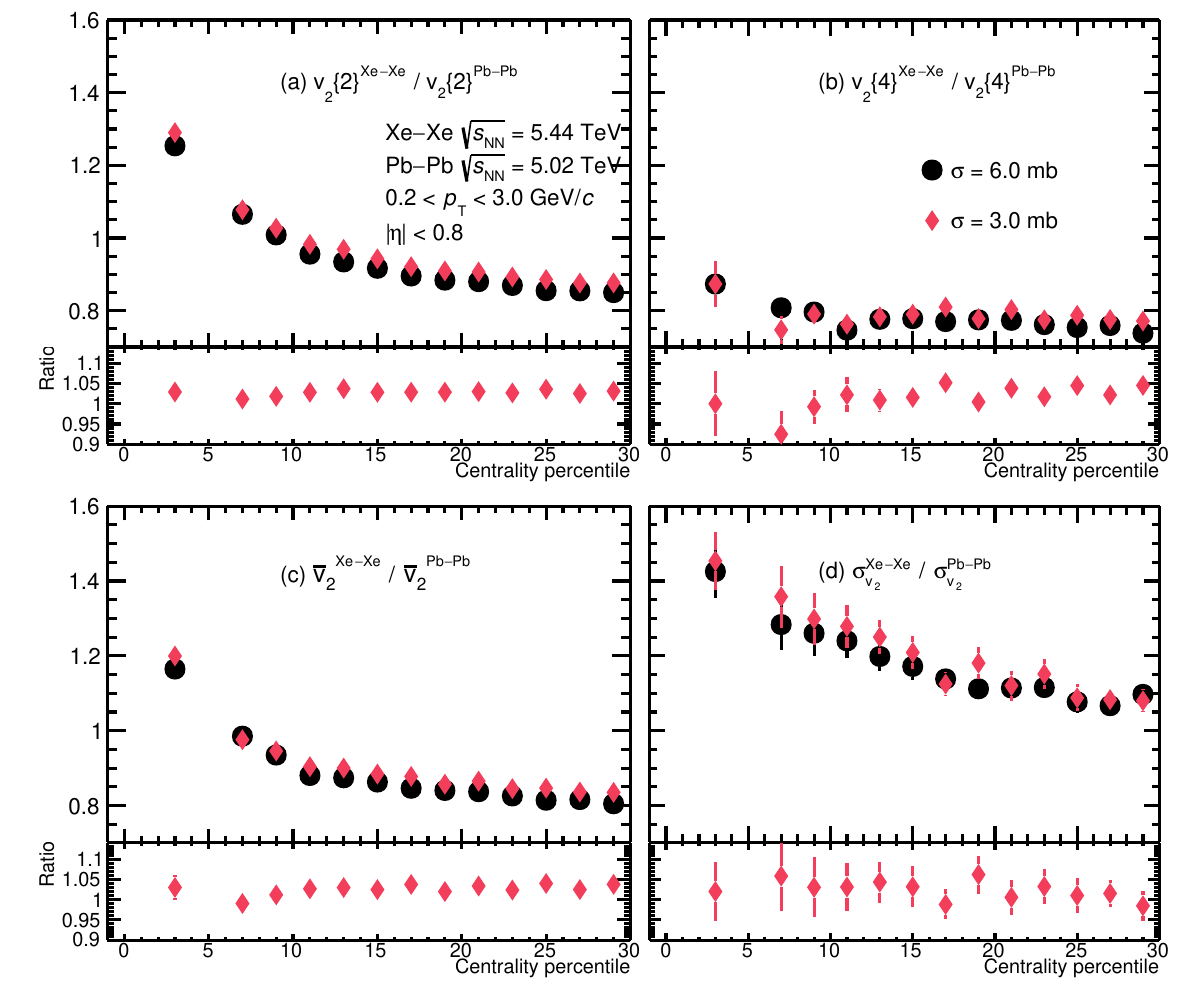}
      \caption[.]{Centrality dependence of $v_2\{2\}$, $v_2\{4\}$, $\bar{v}_{2}$, $\sigma_{v_2}$ in Xe--Xe, Pb--Pb collisions at $\sqrt{s_\mathrm{NN}}$ = 5.44 TeV and $\sqrt{s_\mathrm{NN}}$ = 5.02 TeV respectively in AMPT. The lower panels also present the ratio of different partonic cross sections.}
      \label{FigureMuOverPb_Collect0}
    \end{center}
\end{figure}

In Fig.~\ref{FigureMuOverPb_Collect0}, the centrality dependence of $v_2\{2\}$, $v_2\{4\}$, $\bar{v}_{2}$ and $\sigma_{v_2}$ in the ratio of Xe--Xe/Pb--Pb collisions are presented, using two different partonic cross sections.
The bottom panels of each figure show the ratio of the results using 3 mb and 6 mb (``double ratio").
The ratios in Fig.~\ref{FigureMuOverPb_Collect0} are all consistent with unity, indicating that these observables, including $v_2\{2\}$, $v_2\{4\}$, $\bar{v}_{2}$, $\sigma_{v_2}$, in the ratio of Xe--Xe/Pb--Pb are less dependent on the dynamic evolution and potentially are good probes to initial conditions.

\begin{figure}[!htb]
    \begin{center}
      \includegraphics[width=\textwidth]{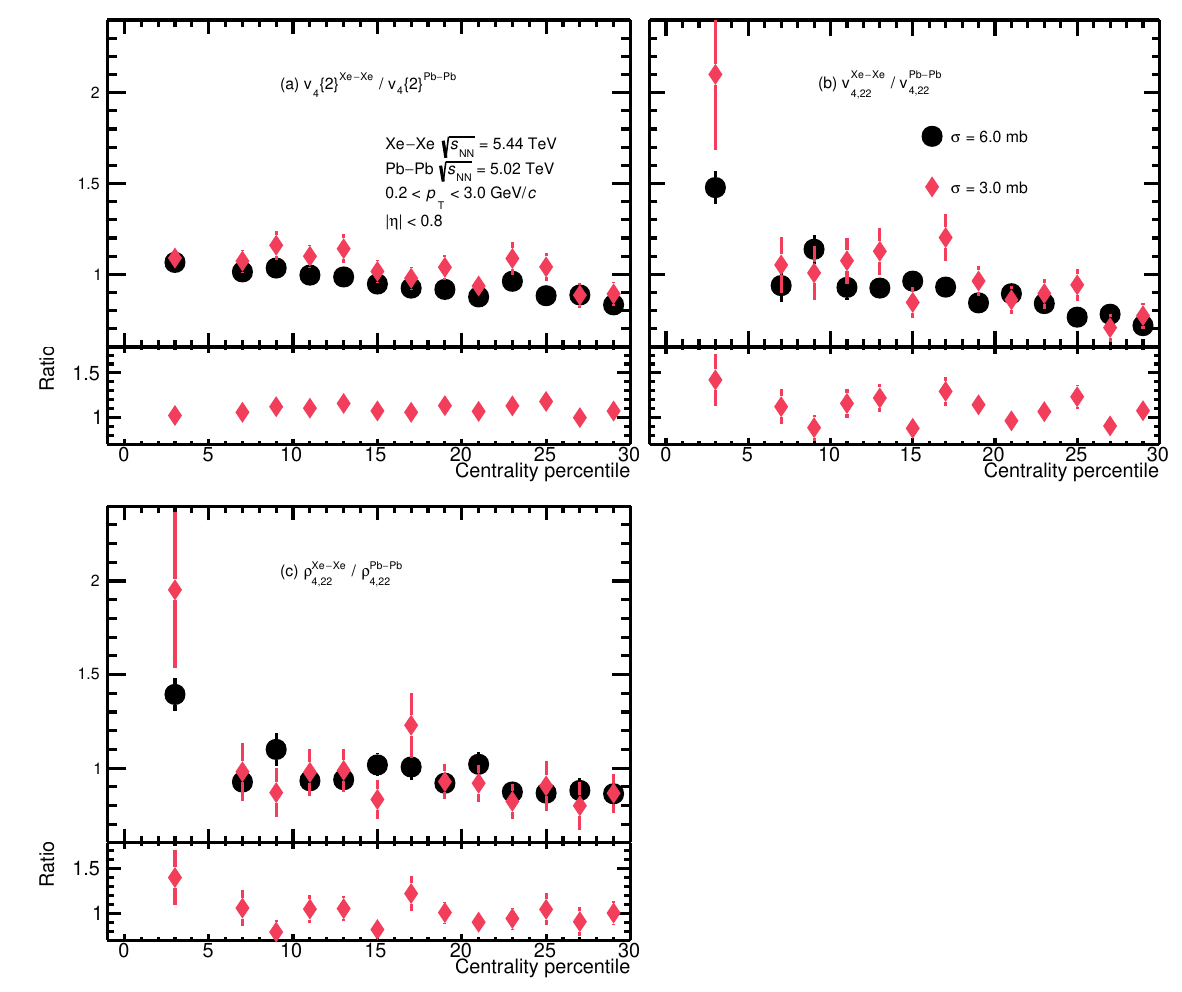}
      \caption[.]{Centrality dependence of $v_4\{2\}$, $v_{4,22}$, $\rho_{4,22}$ in Xe--Xe, Pb--Pb collisions at $\sqrt{s_\mathrm{NN}}$ = 5.44 TeV and $\sqrt{s_\mathrm{NN}}$ = 5.02 TeV respectively in AMPT. The ratio of different partonic cross sections is also presented in the lower panels.}
      \label{FigureMuOverPb_Collect1}
    \end{center}
\end{figure}

In addition, $v_4\{2\}$ and its nonlinear modes $v_{4,22}$ and $\rho_{4,22}$ in the ratio of Xe--Xe/Pb--Pb are also investigated, the centrality dependence is presented in Fig.~\ref{FigureMuOverPb_Collect1}.
Despite large uncertainties, the results with different partonic cross sections are consistent with each other, and the ``double ratio" is compatible with unity.
Such results show that $v_4\{2\}$, $v_{4,22}$, $\rho_{4,22}$ in the ratio of Xe--Xe/Pb--Pb might be not affected by dynamic evolution.
Considering $v_4\{2\}$ and $v_{4,22}$ are sensitive to $a_0$ in the mid-central region, while $v_{4,22}$ and $\rho_{4,22}$ are enhanced by non-zero $\beta_2$ in the most central region, these observables are valuable in the exploration of initial conditions and nuclear structure parameters.

\section{Summary}
\label{sec:4}

This paper presents a comprehensive exploration of the influence of nuclear deformation and nuclear diffuseness on various flow observables in Xe--Xe collisions at $\sqrt{s_\mathrm{NN}}$ = 5.44 TeV, using AMPT event generator. We observe that the elliptic flow coefficients $v_{2}\{2\}$, $\bar{v}_{2}$, elliptic flow fluctuation $\sigma_{v_2}$, as well as the nonlinear flow mode observables $v_{4,22}$ and $\rho_{4,22}$, are enhanced in central collisions in the presence of nuclear quadrupole deformation $\beta_2$. These enhancements come from the increased eccentricities and their fluctuations in the initial geometry in the presence of a deformed shape of $^{129}$Xe. Thus, the aforementioned flow observables can potentially constrain nuclear deformation through data-model comparisons in high-energy heavy-ion collisions in the near future. On the other hand, most flow observables do not exhibit much sensitivity to the triaxiality parameter, underscoring the importance of utilizing combined flow observables like $v_n-[p_{\rm T}]$ correlations in future studies.

%from their only sensitivity to $a_0$ in this study
In addition, larger nuclear diffuseness $a_0$ for $^{129}$Xe leads to smaller values in $v_{2}\{2\}$, $\bar{v}_2$, $v_{4}\{2\}$, $v_{4,22}$, and $v_2$ with multi-particle correlations~($v_{2}\{4\},v_{2}\{6\},v_{2}\{8\}$) in mid-central collisions. Among them, $v_{4}\{2\}$, $v_{2}\{4\}$, $v_{2}\{6\}$, and $v_{2}\{8\}$ are only sensitive to $a_0$, while $\sigma_{v_2}$ and $\rho_{4,22}$ are only sensitive to $\beta_2$. The differing sensitivities of these observables allow for separating constraints on the values of $a_0$ and $\beta_2$, which will help fine-tune other model parameters and enable more precise predictions. 

We also investigate the influence of dynamic evolution on flow observables by varying the parton cross-section in the AMPT model. The parton cross-section does not significantly affect most of the observables in the ratio of Xe--Xe/Pb--Pb, indicating that the final state effect is canceled out when comparing the two collision systems. Future collisions of different nuclear species at varying energies will provide more insights into heavy ion collisions and improve our nuclear structure knowledge. 

\section{Acknowledgements}
\label{sec:5}

M. Zhao and Y.Zhou are funded by the European Union (ERC, InitialConditions), VILLUM FONDEN (grant number 00025462), and Danmarks Frie Forskningsfond (Independent Research Fund Denmark).  J. Jia's work is supported by the US Department of Energy (grant number DE-FG02-87ER40331).

\section{Appendix}
\label{sec:6}
\subsection{nuclear deforamtion}
\label{sec:6:1}
\begin{figure}[!htb]
    \begin{center}
      \includegraphics[width=\textwidth]{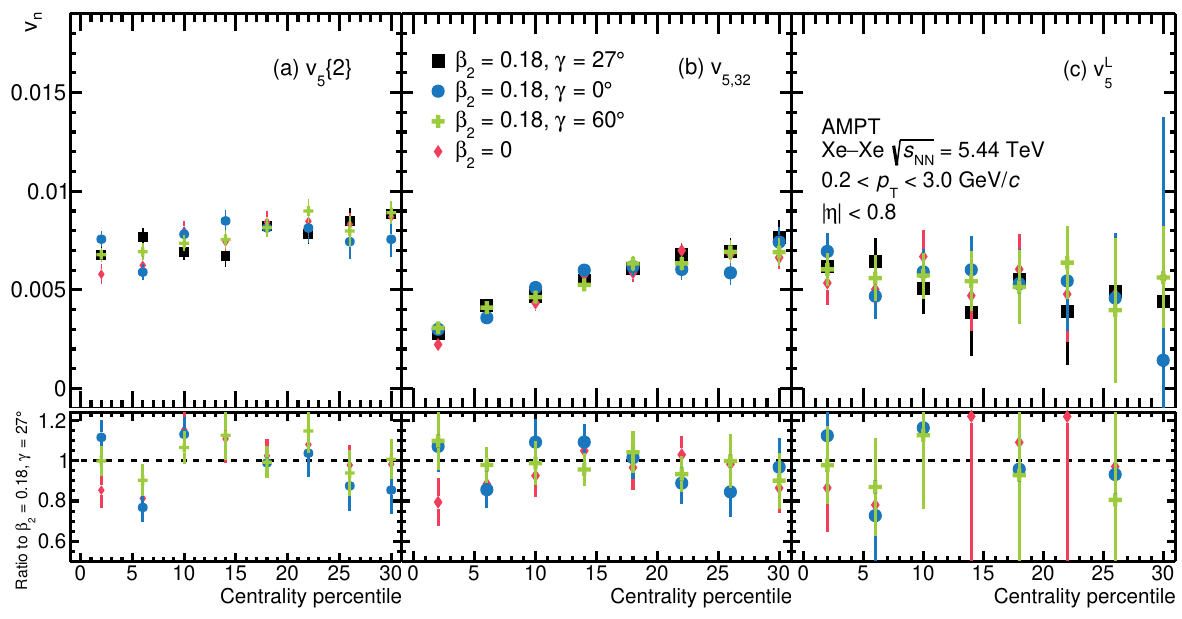}
      \caption[.]{Centrality dependence of nonlinear modes including (a)$v_{5}$, (b)$v_{5,32}$ and (c)Linear $v_{5}$ in Xe--Xe collisions at $\sqrt{s_\mathrm{NN}}$ = 5.44 TeV in AMPT.}
      \label{FigureNonlinearVn5}
    \end{center}
\end{figure}
\begin{figure}[!htb]
    \begin{center}
      \includegraphics[width=\textwidth]{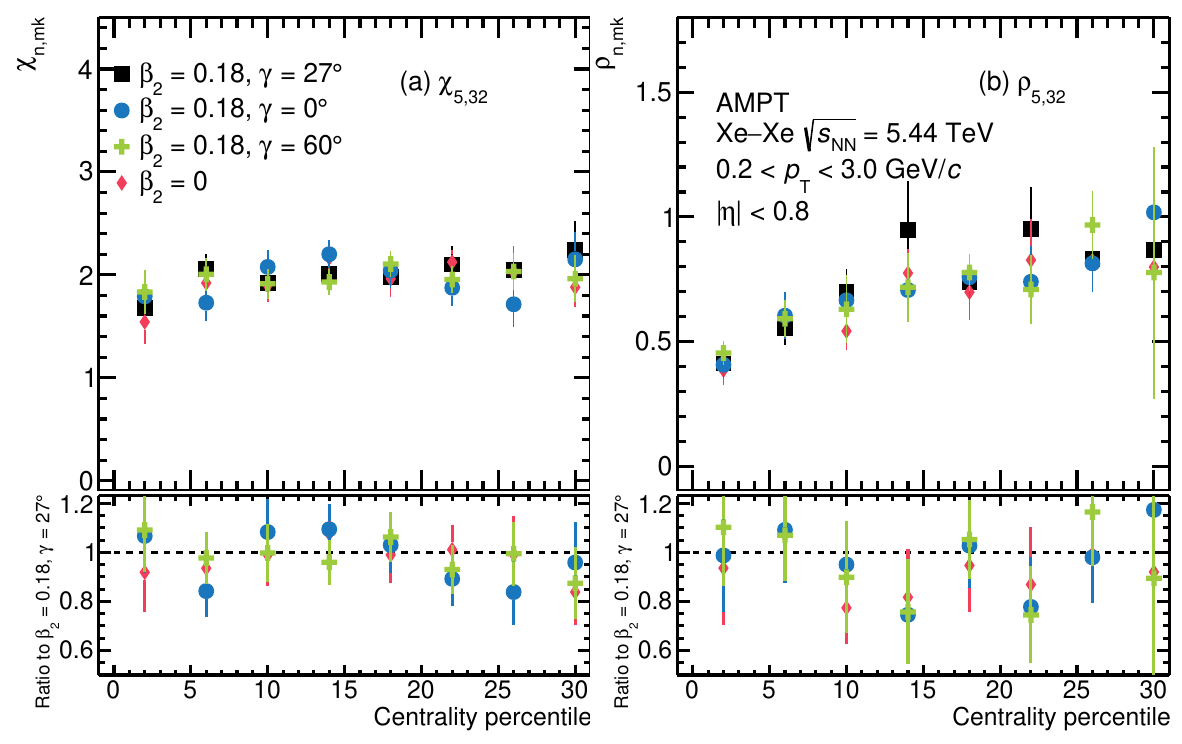}
      \caption[.]{Centrality dependence of nonlinear modes including (a)$\chi_{5,32}$ and (b)$\rho_{5,32}$ in Xe--Xe collisions at $\sqrt{s_\mathrm{NN}}$ = 5.44 TeV in AMPT.}
      \label{FigureNonlinearMode5}
    \end{center}
\end{figure}
\clearpage
\subsection{nuclear diffuseness}
\label{sec:6:2}
\begin{figure}[!htb]
    \begin{center}
      \includegraphics[width=\textwidth]{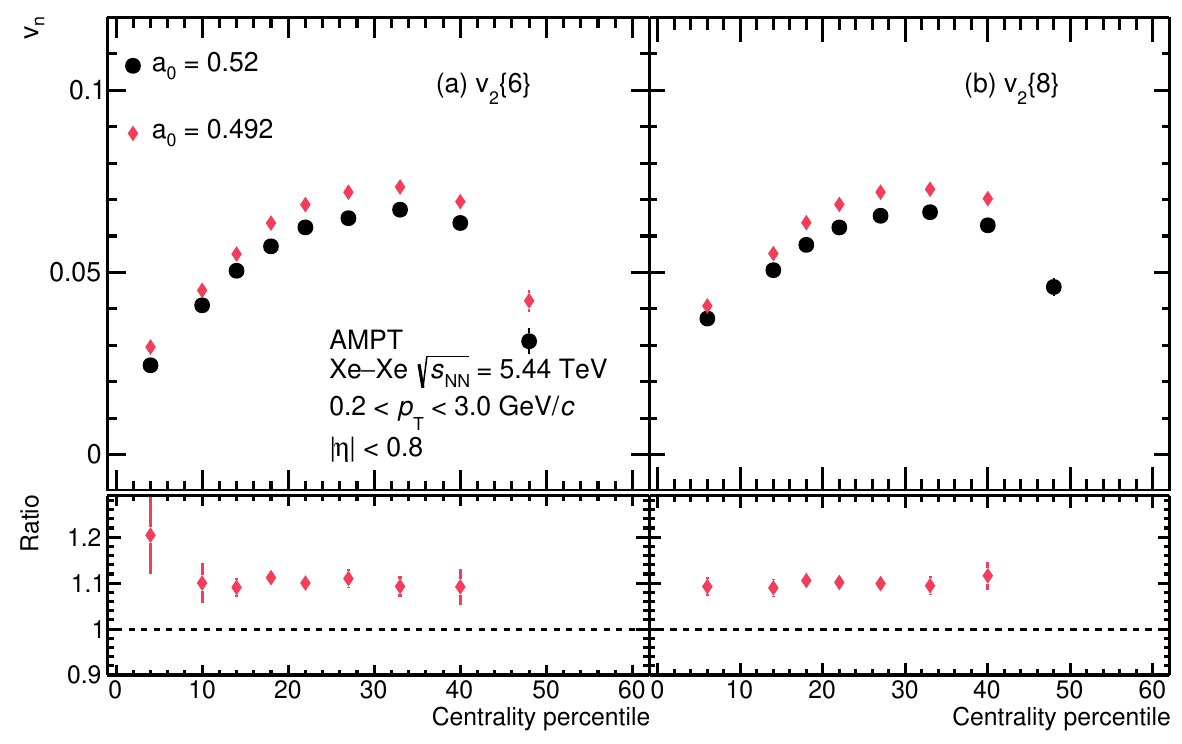}
      \caption[.]{Centrality dependence of $v_2\{6\},v_2\{8\}$ in Xe--Xe collisions at $\sqrt{s_\mathrm{NN}}$ = 5.44 TeV in AMPT.}
      \label{FigureA0Setv26v28}
    \end{center}
  \end{figure}
  \begin{figure}[!htb]
    \begin{center}
      \includegraphics[width=\textwidth]{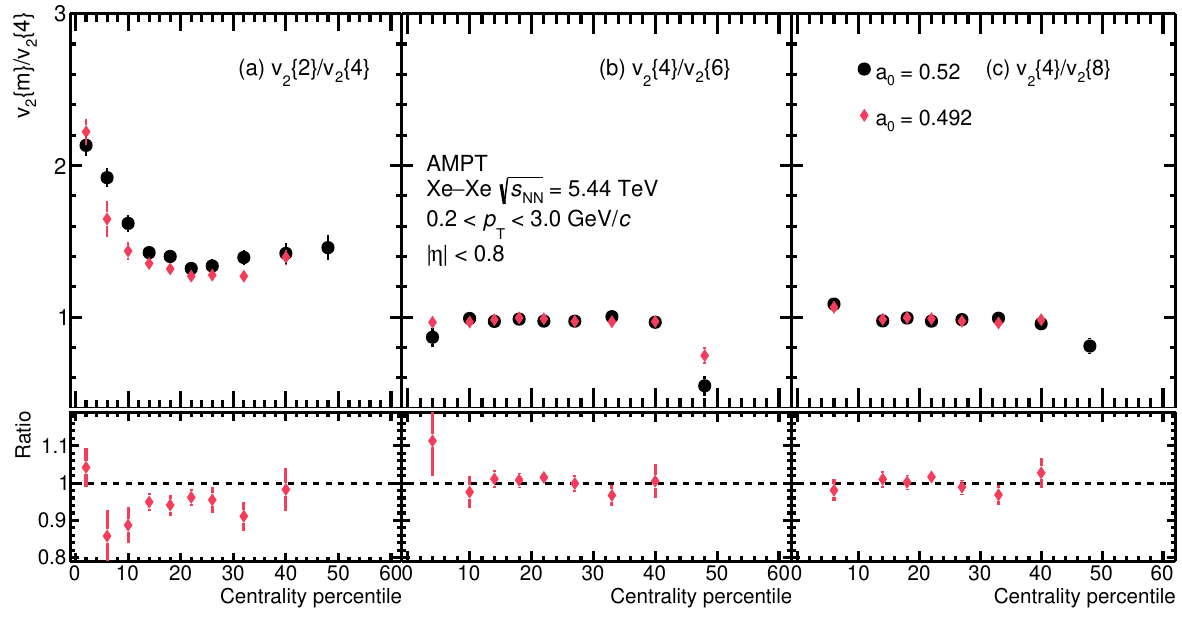}
      \caption[.]{Centrality dependence of $v_2\{2\}/v_2\{4\}$,$v_2\{6\}/v_2\{4\}$,$v_2\{8\}/v_2\{4\}$ in Xe--Xe collisions at $\sqrt{s_\mathrm{NN}}$ = 5.44 TeV in AMPT.}
      \label{FigureA0v22v26v28overv24}
    \end{center}
  \end{figure}
  \begin{figure}[!htb]
    \begin{center}
      \includegraphics[width=\textwidth]{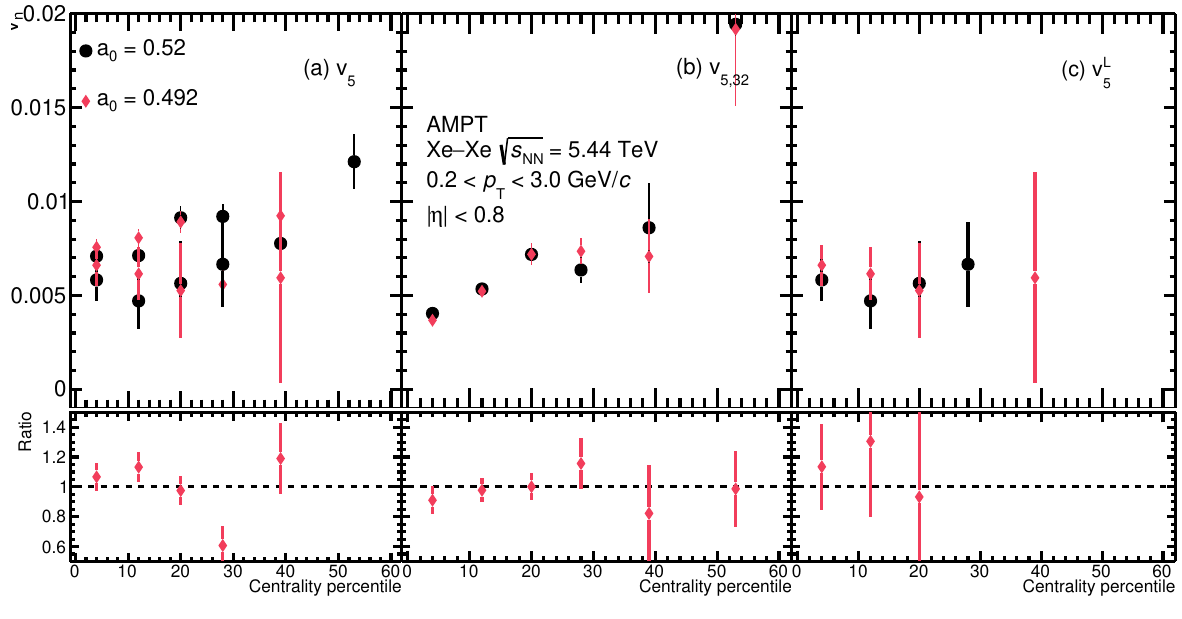}
      \caption[.]{Centrality dependence of nonlinear modes including (a)$v_{4}$, (b)$v_{4,22}$ and (c)Linear $v_{4}$ in Xe--Xe collisions at $\sqrt{s_\mathrm{NN}}$ = 5.44 TeV in AMPT.}
      \label{FigureA0NonlinearVn5}
    \end{center}
\end{figure}
\begin{figure}[!htb]
    \begin{center}
      \includegraphics[width=\textwidth]{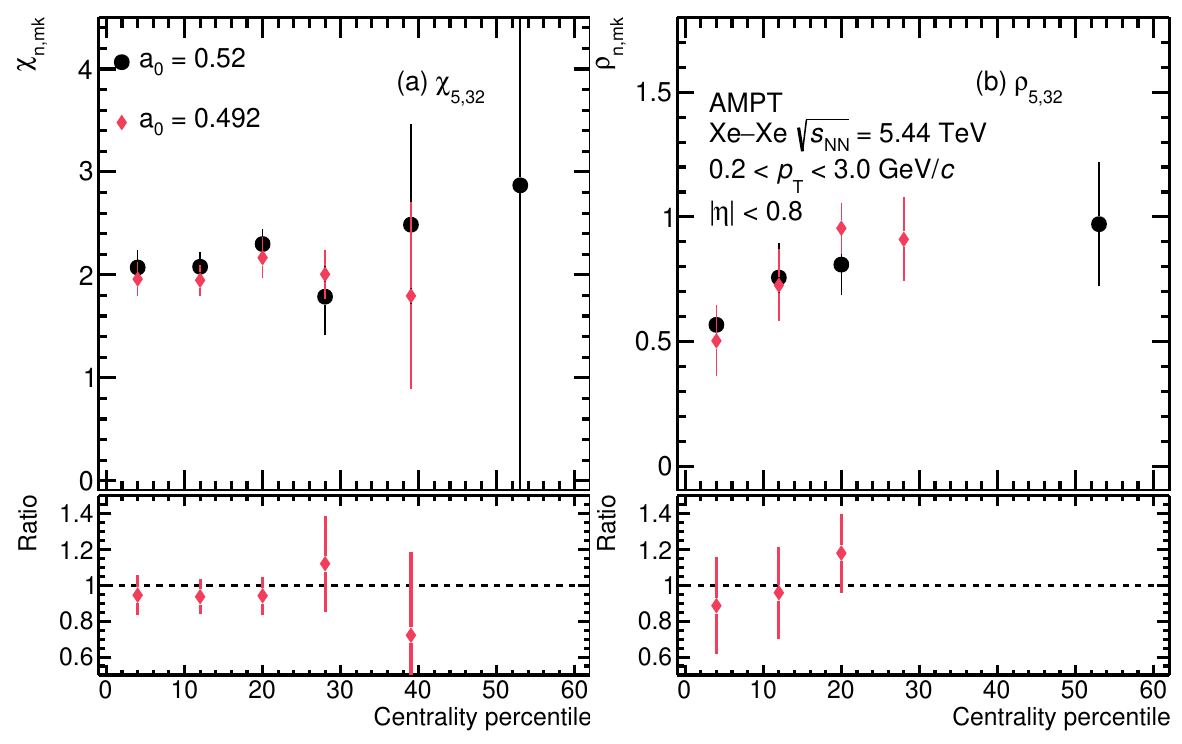}
      \caption[.]{Centrality dependence of nonlinear modes including (a)$\chi_{5,32}$ and (b)$\rho_{5,32}$ in Xe--Xe collisions at $\sqrt{s_\mathrm{NN}}$ = 5.44 TeV in AMPT.}
      \label{FigureA0Rho}
    \end{center}
\end{figure}
\begin{figure}[!htb]
    \begin{center}
      \includegraphics[width=0.8\textwidth]{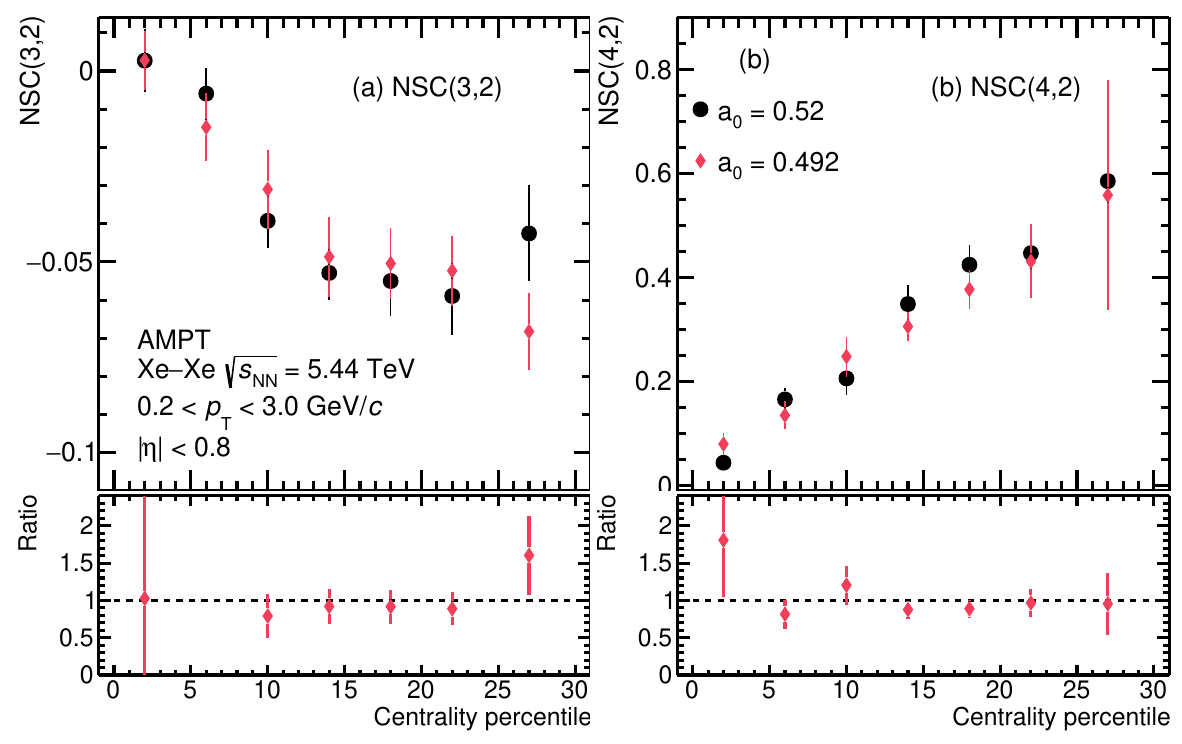}
      \caption[.]{Centrality dependence of NSC$(m,n)$ in Xe--Xe collisions at $\sqrt{s_\mathrm{NN}}$ = 5.44 TeV in AMPT.}
      \label{FigureA0NSC}
    \end{center}
  \end{figure}
\bibliographystyle{utphys}
\bibliography{Reference}

\providecommand{\href}[2]{#2}\begingroup\raggedright\begin{thebibliography}{10}

\bibitem{Shuryak:1980tp}
E.~V. Shuryak, ``{Quantum Chromodynamics and the Theory of Superdense
  Matter},'' \href{http://dx.doi.org/10.1016/0370-1573(80)90105-2}{{\em Phys.
  Rept.} {\bfseries 61} (1980) 71--158}.

\bibitem{Shuryak:1978ij}
E.~V. Shuryak, ``{Quark-Gluon Plasma and Hadronic Production of Leptons,
  Photons and Psions},''
  \href{http://dx.doi.org/10.1016/0370-2693(78)90370-2}{{\em Phys. Lett. B}
  {\bfseries 78} (1978) 150}.

\bibitem{Lacey:2006bc}
R.~A. Lacey, N.~N. Ajitanand, J.~M. Alexander, P.~Chung, W.~G. Holzmann,
  M.~Issah, A.~Taranenko, P.~Danielewicz, and H.~Stoecker, ``{Has the QCD
  Critical Point been Signaled by Observations at RHIC?},''
  \href{http://dx.doi.org/10.1103/PhysRevLett.98.092301}{{\em Phys. Rev. Lett.}
  {\bfseries 98} (2007) 092301},
  \href{http://arxiv.org/abs/nucl-ex/0609025}{{\ttfamily
  arXiv:nucl-ex/0609025}}.

\bibitem{Molnar:2008xj}
D.~Molnar and P.~Huovinen, ``{Dissipative effects from transport and viscous
  hydrodynamics},''
  \href{http://dx.doi.org/10.1088/0954-3899/35/10/104125}{{\em J. Phys. G}
  {\bfseries 35} (2008) 104125},
  \href{http://arxiv.org/abs/0806.1367}{{\ttfamily arXiv:0806.1367 [nucl-th]}}.

\bibitem{Muller:2012zq}
B.~Muller, J.~Schukraft, and B.~Wyslouch, ``{First Results from Pb+Pb
  collisions at the LHC},''
  \href{http://dx.doi.org/10.1146/annurev-nucl-102711-094910}{{\em Ann. Rev.
  Nucl. Part. Sci.} {\bfseries 62} (2012) 361--386},
  \href{http://arxiv.org/abs/1202.3233}{{\ttfamily arXiv:1202.3233 [hep-ex]}}.

\bibitem{Drescher:2007cd}
H.-J. Drescher, A.~Dumitru, C.~Gombeaud, and J.-Y. Ollitrault, ``{The
  Centrality dependence of elliptic flow, the hydrodynamic limit, and the
  viscosity of hot QCD},''
  \href{http://dx.doi.org/10.1103/PhysRevC.76.024905}{{\em Phys. Rev. C}
  {\bfseries 76} (2007) 024905},
  \href{http://arxiv.org/abs/0704.3553}{{\ttfamily arXiv:0704.3553 [nucl-th]}}.

\bibitem{Heinz:2013th}
U.~Heinz and R.~Snellings, ``{Collective flow and viscosity in relativistic
  heavy-ion collisions},''
  \href{http://dx.doi.org/10.1146/annurev-nucl-102212-170540}{{\em Ann. Rev.
  Nucl. Part. Sci.} {\bfseries 63} (2013) 123--151},
  \href{http://arxiv.org/abs/1301.2826}{{\ttfamily arXiv:1301.2826 [nucl-th]}}.

\bibitem{Molnar:2001ux}
D.~Molnar and M.~Gyulassy, ``{Saturation of elliptic flow and the transport
  opacity of the gluon plasma at RHIC},''
  \href{http://dx.doi.org/10.1016/S0375-9474(01)01224-6}{{\em Nucl. Phys. A}
  {\bfseries 697} (2002) 495--520},
  \href{http://arxiv.org/abs/nucl-th/0104073}{{\ttfamily
  arXiv:nucl-th/0104073}}. [Erratum: Nucl.Phys.A 703, 893--894 (2002)].

\bibitem{Song:2017wtw}
H.~Song, Y.~Zhou, and K.~Gajdosova, ``{Collective flow and hydrodynamics in
  large and small systems at the LHC},''
  \href{http://dx.doi.org/10.1007/s41365-017-0245-4}{{\em Nucl. Sci. Tech.}
  {\bfseries 28} no.~7, (2017) 99},
  \href{http://arxiv.org/abs/1703.00670}{{\ttfamily arXiv:1703.00670
  [nucl-th]}}.

\bibitem{Teaney:2003kp}
D.~Teaney, ``{The Effects of viscosity on spectra, elliptic flow, and HBT
  radii},'' \href{http://dx.doi.org/10.1103/PhysRevC.68.034913}{{\em Phys. Rev.
  C} {\bfseries 68} (2003) 034913},
  \href{http://arxiv.org/abs/nucl-th/0301099}{{\ttfamily
  arXiv:nucl-th/0301099}}.

\bibitem{Xu:2007jv}
Z.~Xu, C.~Greiner, and H.~Stocker, ``{PQCD calculations of elliptic flow and
  shear viscosity at RHIC},''
  \href{http://dx.doi.org/10.1103/PhysRevLett.101.082302}{{\em Phys. Rev.
  Lett.} {\bfseries 101} (2008) 082302},
  \href{http://arxiv.org/abs/0711.0961}{{\ttfamily arXiv:0711.0961 [nucl-th]}}.

\bibitem{Voloshin:1994mz}
S.~Voloshin and Y.~Zhang, ``{Flow study in relativistic nuclear collisions by
  Fourier expansion of Azimuthal particle distributions},''
  \href{http://dx.doi.org/10.1007/s002880050141}{{\em Z. Phys. C} {\bfseries
  70} (1996) 665--672}, \href{http://arxiv.org/abs/hep-ph/9407282}{{\ttfamily
  arXiv:hep-ph/9407282}}.

\bibitem{Niemi:2012aj}
H.~Niemi, G.~S. Denicol, H.~Holopainen, and P.~Huovinen, ``{Event-by-event
  distributions of azimuthal asymmetries in ultrarelativistic heavy-ion
  collisions},'' \href{http://dx.doi.org/10.1103/PhysRevC.87.054901}{{\em Phys.
  Rev. C} {\bfseries 87} no.~5, (2013) 054901},
  \href{http://arxiv.org/abs/1212.1008}{{\ttfamily arXiv:1212.1008 [nucl-th]}}.

\bibitem{Bilandzic:2013kga}
A.~Bilandzic, C.~H. Christensen, K.~Gulbrandsen, A.~Hansen, and Y.~Zhou,
  ``{Generic framework for anisotropic flow analyses with multiparticle
  azimuthal correlations},''
  \href{http://dx.doi.org/10.1103/PhysRevC.89.064904}{{\em Phys. Rev. C}
  {\bfseries 89} no.~6, (2014) 064904},
  \href{http://arxiv.org/abs/1312.3572}{{\ttfamily arXiv:1312.3572 [nucl-ex]}}.

\bibitem{ATLAS:2015qwl}
{\bfseries ATLAS} Collaboration, G.~Aad {\em et~al.}, ``{Measurement of the
  correlation between flow harmonics of different order in lead-lead collisions
  at $\sqrt{s_{NN}}$=2.76 TeV with the ATLAS detector},''
  \href{http://dx.doi.org/10.1103/PhysRevC.92.034903}{{\em Phys. Rev. C}
  {\bfseries 92} no.~3, (2015) 034903},
  \href{http://arxiv.org/abs/1504.01289}{{\ttfamily arXiv:1504.01289
  [hep-ex]}}.

\bibitem{Qian:2016pau}
J.~Qian and U.~Heinz, ``{Hydrodynamic flow amplitude correlations in
  event-by-event fluctuating heavy-ion collisions},''
  \href{http://dx.doi.org/10.1103/PhysRevC.94.024910}{{\em Phys. Rev. C}
  {\bfseries 94} no.~2, (2016) 024910},
  \href{http://arxiv.org/abs/1607.01732}{{\ttfamily arXiv:1607.01732
  [nucl-th]}}.

\bibitem{Zhu:2016puf}
X.~Zhu, Y.~Zhou, H.~Xu, and H.~Song, ``{Correlations of flow harmonics in 2.76A
  TeV Pb--Pb collisions},''
  \href{http://dx.doi.org/10.1103/PhysRevC.95.044902}{{\em Phys. Rev. C}
  {\bfseries 95} no.~4, (2017) 044902},
  \href{http://arxiv.org/abs/1608.05305}{{\ttfamily arXiv:1608.05305
  [nucl-th]}}.

\bibitem{Bernhard:2019bmu}
J.~E. Bernhard, J.~S. Moreland, and S.~A. Bass, ``{Bayesian estimation of the
  specific shear and bulk viscosity of quark\textendash{}gluon plasma},''
  \href{http://dx.doi.org/10.1038/s41567-019-0611-8}{{\em Nature Phys.}
  {\bfseries 15} no.~11, (2019) 1113--1117}.

\bibitem{JETSCAPE:2020mzn}
{\bfseries JETSCAPE} Collaboration, D.~Everett {\em et~al.}, ``{Multisystem
  Bayesian constraints on the transport coefficients of QCD matter},''
  \href{http://dx.doi.org/10.1103/PhysRevC.103.054904}{{\em Phys. Rev. C}
  {\bfseries 103} no.~5, (2021) 054904},
  \href{http://arxiv.org/abs/2011.01430}{{\ttfamily arXiv:2011.01430
  [hep-ph]}}.

\bibitem{Nijs:2020ors}
G.~Nijs, W.~van~der Schee, U.~G\"ursoy, and R.~Snellings, ``{Transverse
  Momentum Differential Global Analysis of Heavy-Ion Collisions},''
  \href{http://dx.doi.org/10.1103/PhysRevLett.126.202301}{{\em Phys. Rev.
  Lett.} {\bfseries 126} no.~20, (2021) 202301},
  \href{http://arxiv.org/abs/2010.15130}{{\ttfamily arXiv:2010.15130
  [nucl-th]}}.

\bibitem{Parkkila:2021tqq}
J.~E. Parkkila, A.~Onnerstad, and D.~J. Kim, ``{Bayesian estimation of the
  specific shear and bulk viscosity of the quark-gluon plasma with additional
  flow harmonic observables},''
  \href{http://dx.doi.org/10.1103/PhysRevC.104.054904}{{\em Phys. Rev. C}
  {\bfseries 104} no.~5, (2021) 054904},
  \href{http://arxiv.org/abs/2106.05019}{{\ttfamily arXiv:2106.05019
  [hep-ph]}}.

\bibitem{Jia:2021tzt}
J.~Jia, ``{Shape of atomic nuclei in heavy ion collisions},''
  \href{http://dx.doi.org/10.1103/PhysRevC.105.014905}{{\em Phys. Rev. C}
  {\bfseries 105} no.~1, (2022) 014905},
  \href{http://arxiv.org/abs/2106.08768}{{\ttfamily arXiv:2106.08768
  [nucl-th]}}.

\bibitem{Zhang:2021kxj}
C.~Zhang and J.~Jia, ``{Evidence of Quadrupole and Octupole Deformations in
  Zr96+Zr96 and Ru96+Ru96 Collisions at Ultrarelativistic Energies},''
  \href{http://dx.doi.org/10.1103/PhysRevLett.128.022301}{{\em Phys. Rev.
  Lett.} {\bfseries 128} no.~2, (2022) 022301},
  \href{http://arxiv.org/abs/2109.01631}{{\ttfamily arXiv:2109.01631
  [nucl-th]}}.

\bibitem{Giacalone:2021udy}
G.~Giacalone, J.~Jia, and C.~Zhang, ``Impact of nuclear deformation on
  relativistic heavy-ion collisions: Assessing consistency in nuclear physics
  across energy scales,''.
  \url{https://link.aps.org/doi/10.1103/PhysRevLett.127.242301}.

\bibitem{Jia:2022qgl}
J.~Jia, G.~Giacalone, and C.~Zhang, ``{Separating the impact of nuclear skin
  and nuclear deformation on elliptic flow and its fluctuations in high-energy
  isobar collisions},'' \href{http://arxiv.org/abs/2206.10449}{{\ttfamily
  arXiv:2206.10449 [nucl-th]}}.

\bibitem{Magdy:2022cvt}
N.~Magdy, ``{Impact of nuclear deformation on collective flow observables in
  relativistic U+U collisions},''
  \href{http://dx.doi.org/10.1140/epja/s10050-023-00982-0}{{\em Eur. Phys. J.
  A} {\bfseries 59} no.~3, (2023) 64},
  \href{http://arxiv.org/abs/2206.05332}{{\ttfamily arXiv:2206.05332
  [nucl-th]}}.

\bibitem{Jia:2022qrq}
J.~Jia, G.~Giacalone, and C.~Zhang, ``{Precision Tests of the Nonlinear Mode
  Coupling of Anisotropic Flow via High-Energy Collisions of Isobars},''
  \href{http://dx.doi.org/10.1088/0256-307X/40/4/042501}{{\em Chin. Phys.
  Lett.} {\bfseries 40} no.~4, (2023) 042501},
  \href{http://arxiv.org/abs/2206.07184}{{\ttfamily arXiv:2206.07184
  [nucl-th]}}.

\bibitem{Xu:2021uar}
H.-j. Xu, W.~Zhao, H.~Li, Y.~Zhou, L.-W. Chen, and F.~Wang, ``{Probing nuclear
  structure with mean transverse momentum in relativistic isobar collisions},''
  \href{http://arxiv.org/abs/2111.14812}{{\ttfamily arXiv:2111.14812
  [nucl-th]}}.

\bibitem{Jia:2021qyu}
J.~Jia, ``{Probing triaxial deformation of atomic nuclei in high-energy heavy
  ion collisions},'' \href{http://dx.doi.org/10.1103/PhysRevC.105.044905}{{\em
  Phys. Rev. C} {\bfseries 105} no.~4, (2022) 044905},
  \href{http://arxiv.org/abs/2109.00604}{{\ttfamily arXiv:2109.00604
  [nucl-th]}}.

\bibitem{ALICE:2018yvr}
{\bfseries ALICE} Collaboration, ``{Centrality determination using the Glauber
  model in Xe-Xe collisions at $\sqrt{s_{\rm NN}} = 5.44$ TeV},''.

\bibitem{Tsukada:2017llu}
K.~Tsukada {\em et~al.}, ``{First elastic electron scattering from $^{132}$Xe
  at the SCRIT facility},''
  \href{http://dx.doi.org/10.1103/PhysRevLett.118.262501}{{\em Phys. Rev.
  Lett.} {\bfseries 118} no.~26, (2017) 262501},
  \href{http://arxiv.org/abs/1703.04278}{{\ttfamily arXiv:1703.04278
  [nucl-ex]}}.

\bibitem{Fischer:1974aaa}
W.~Fischer {\em et~al.}, ``{Isotope shifts in the atomic spectrum of xenon and
  nuclear deformation effects},''
  \href{http://dx.doi.org/10.1007/BF01677442}{{\em Zeitschrift für Physik}
  {\bfseries 270} (1974) 113–120}.

\bibitem{Kumar:1972zza}
K.~Kumar, ``{Intrinsic Quadrupole Moments and Shapes of Nuclear Ground States
  and Excited States},''
  \href{http://dx.doi.org/10.1103/PhysRevLett.28.249}{{\em Phys. Rev. Lett.}
  {\bfseries 28} (1972) 249--253}.

\bibitem{Poves:2019byh}
A.~Poves, F.~Nowacki, and Y.~Alhassid, ``{Limits on assigning a shape to a
  nucleus},'' \href{http://dx.doi.org/10.1103/PhysRevC.101.054307}{{\em Phys.
  Rev. C} {\bfseries 101} no.~5, (2020) 054307},
  \href{http://arxiv.org/abs/1906.07542}{{\ttfamily arXiv:1906.07542
  [nucl-th]}}.

\bibitem{Cline:1986ik}
D.~Cline, ``{Nuclear shapes studied by coulomb excitation},''
  \href{http://dx.doi.org/10.1146/annurev.ns.36.120186.003343}{{\em Ann. Rev.
  Nucl. Part. Sci.} {\bfseries 36} (1986) 683--716}.

\bibitem{Morrison:2020azy}
L.~Morrison {\em et~al.}, ``{Quadrupole deformation of $^{130}$Xe measured in a
  Coulomb-excitation experiment},''
  \href{http://dx.doi.org/10.1103/PhysRevC.102.054304}{{\em Phys. Rev. C}
  {\bfseries 102} no.~5, (2020) 054304}.

\bibitem{ALICE:2018lao}
{\bfseries ALICE} Collaboration, S.~Acharya {\em et~al.}, ``{Anisotropic flow
  in Xe-Xe collisions at $\mathbf{\sqrt{s_{\rm{NN}}} = 5.44}$ TeV},''
  \href{http://dx.doi.org/10.1016/j.physletb.2018.06.059}{{\em Phys. Lett. B}
  {\bfseries 784} (2018) 82--95},
  \href{http://arxiv.org/abs/1805.01832}{{\ttfamily arXiv:1805.01832
  [nucl-ex]}}.

\bibitem{ALICE:2021gxt}
{\bfseries ALICE} Collaboration, S.~Acharya {\em et~al.}, ``{Characterizing the
  initial conditions of heavy-ion collisions at the LHC with mean transverse
  momentum and anisotropic flow correlations},''
  \href{http://dx.doi.org/10.1016/j.physletb.2022.137393}{{\em Phys. Lett. B}
  {\bfseries 834} (2022) 137393},
  \href{http://arxiv.org/abs/2111.06106}{{\ttfamily arXiv:2111.06106
  [nucl-ex]}}.

\bibitem{Lin:2004en}
Z.-W. Lin, C.~M. Ko, B.-A. Li, B.~Zhang, and S.~Pal, ``{A Multi-phase transport
  model for relativistic heavy ion collisions},''
  \href{http://dx.doi.org/10.1103/PhysRevC.72.064901}{{\em Phys. Rev. C}
  {\bfseries 72} (2005) 064901},
  \href{http://arxiv.org/abs/nucl-th/0411110}{{\ttfamily
  arXiv:nucl-th/0411110}}.

\bibitem{Bhaduri:2010wi}
P.~P. Bhaduri and S.~Chattopadhyay, ``{Differential elliptic flow of identified
  hadrons and constituent quark number scaling at FAIR},''
  \href{http://dx.doi.org/10.1103/PhysRevC.81.034906}{{\em Phys. Rev. C}
  {\bfseries 81} (2010) 034906},
  \href{http://arxiv.org/abs/1002.4100}{{\ttfamily arXiv:1002.4100 [hep-ph]}}.

\bibitem{Guo:2019joy}
Y.~Guo, S.~Shi, S.~Feng, and J.~Liao, ``{Magnetic Field Induced Polarization
  Difference between Hyperons and Anti-hyperons},''
  \href{http://dx.doi.org/10.1016/j.physletb.2019.134929}{{\em Phys. Lett. B}
  {\bfseries 798} (2019) 134929},
  \href{http://arxiv.org/abs/1905.12613}{{\ttfamily arXiv:1905.12613
  [nucl-th]}}.

\bibitem{Haque:2019vgi}
M.~R. Haque, M.~Nasim, and B.~Mohanty, ``{Systematic investigation of azimuthal
  anisotropy in Au+Au and U+U collisions at $\sqrt {s}_{NN}$ = 200 GeV},''
  \href{http://dx.doi.org/10.1088/1361-6471/ab2ba4}{{\em J. Phys. G} {\bfseries
  46} no.~8, (2019) 085104}.

\bibitem{Ma:2016fve}
G.-L. Ma and Z.-W. Lin, ``{Predictions for $\sqrt {s_{NN}}=5.02$ TeV Pb+Pb
  Collisions from a Multi-Phase Transport Model},''
  \href{http://dx.doi.org/10.1103/PhysRevC.93.054911}{{\em Phys. Rev. C}
  {\bfseries 93} no.~5, (2016) 054911},
  \href{http://arxiv.org/abs/1601.08160}{{\ttfamily arXiv:1601.08160
  [nucl-th]}}.

\bibitem{Magdy:2020bhd}
N.~Magdy, O.~Evdokimov, and R.~A. Lacey, ``{A method to test the coupling
  strength of the linear and nonlinear contributions to higher-order flow
  harmonics via Event Shape Engineering},''
  \href{http://dx.doi.org/10.1088/1361-6471/abcb59}{{\em J. Phys. G} {\bfseries
  48} no.~2, (2020) 025101}, \href{http://arxiv.org/abs/2002.04583}{{\ttfamily
  arXiv:2002.04583 [nucl-ex]}}.

\bibitem{Nasim:2010hw}
M.~Nasim, L.~Kumar, P.~K. Netrakanti, and B.~Mohanty, ``{Energy dependence of
  elliptic flow from heavy-ion collision models},''
  \href{http://dx.doi.org/10.1103/PhysRevC.82.054908}{{\em Phys. Rev. C}
  {\bfseries 82} (2010) 054908},
  \href{http://arxiv.org/abs/1010.5196}{{\ttfamily arXiv:1010.5196 [nucl-ex]}}.

\bibitem{Xu:2010du}
J.~Xu and C.~M. Ko, ``{The effect of triangular flow on di-hadron azimuthal
  correlations in relativistic heavy ion collisions},''
  \href{http://dx.doi.org/10.1103/PhysRevC.83.021903}{{\em Phys. Rev. C}
  {\bfseries 83} (2011) 021903},
  \href{http://arxiv.org/abs/1011.3750}{{\ttfamily arXiv:1011.3750 [nucl-th]}}.

\bibitem{Xu:2011fe}
J.~Xu and C.~M. Ko, ``{Triangular flow in heavy ion collisions in a multiphase
  transport model},'' \href{http://dx.doi.org/10.1103/PhysRevC.84.014903}{{\em
  Phys. Rev. C} {\bfseries 84} (2011) 014903},
  \href{http://arxiv.org/abs/1103.5187}{{\ttfamily arXiv:1103.5187 [nucl-th]}}.

\bibitem{Wang:2000bf}
X.-N. Wang and M.~Gyulassy, ``{Energy and centrality dependence of rapidity
  densities at RHIC},''
  \href{http://dx.doi.org/10.1103/PhysRevLett.86.3496}{{\em Phys. Rev. Lett.}
  {\bfseries 86} (2001) 3496--3499},
  \href{http://arxiv.org/abs/nucl-th/0008014}{{\ttfamily
  arXiv:nucl-th/0008014}}.

\bibitem{Zhang:1997ej}
B.~Zhang, ``{ZPC 1.0.1: A Parton cascade for ultrarelativistic heavy ion
  collisions},'' \href{http://dx.doi.org/10.1016/S0010-4655(98)00010-1}{{\em
  Comput. Phys. Commun.} {\bfseries 109} (1998) 193--206},
  \href{http://arxiv.org/abs/nucl-th/9709009}{{\ttfamily
  arXiv:nucl-th/9709009}}.

\bibitem{Chen:2005mr}
L.-W. Chen and C.~M. Ko, ``{System size dependence of elliptic flows in
  relativistic heavy-ion collisions},''
  \href{http://dx.doi.org/10.1016/j.physletb.2006.01.037}{{\em Phys. Lett. B}
  {\bfseries 634} (2006) 205--209},
  \href{http://arxiv.org/abs/nucl-th/0505044}{{\ttfamily
  arXiv:nucl-th/0505044}}.

\bibitem{Li:1995pra}
B.-A. Li and C.~M. Ko, ``{Formation of superdense hadronic matter in
  high-energy heavy ion collisions},''
  \href{http://dx.doi.org/10.1103/PhysRevC.52.2037}{{\em Phys. Rev. C}
  {\bfseries 52} (1995) 2037--2063},
  \href{http://arxiv.org/abs/nucl-th/9505016}{{\ttfamily
  arXiv:nucl-th/9505016}}.

\bibitem{Bally:2021qys}
B.~Bally, M.~Bender, G.~Giacalone, and V.~Som\`a, ``{Evidence of the triaxial
  structure of $\boldsymbol{^{129}}$Xe at the Large Hadron Collider},''
  \href{http://dx.doi.org/10.1103/PhysRevLett.128.082301}{{\em Phys. Rev.
  Lett.} {\bfseries 128} no.~8, (2022) 082301},
  \href{http://arxiv.org/abs/2108.09578}{{\ttfamily arXiv:2108.09578
  [nucl-th]}}.

\bibitem{Ma:2014pva}
G.-L. Ma and A.~Bzdak, ``{Long-range azimuthal correlations in
  proton\textendash{}proton and proton\textendash{}nucleus collisions from the
  incoherent scattering of partons},''
  \href{http://dx.doi.org/10.1016/j.physletb.2014.10.066}{{\em Phys. Lett. B}
  {\bfseries 739} (2014) 209--213},
  \href{http://arxiv.org/abs/1404.4129}{{\ttfamily arXiv:1404.4129 [hep-ph]}}.

\bibitem{Bzdak:2014dia}
A.~Bzdak and G.-L. Ma, ``{Elliptic and triangular flow in $p$+Pb and peripheral
  Pb+Pb collisions from parton scatterings},''
  \href{http://dx.doi.org/10.1103/PhysRevLett.113.252301}{{\em Phys. Rev.
  Lett.} {\bfseries 113} no.~25, (2014) 252301},
  \href{http://arxiv.org/abs/1406.2804}{{\ttfamily arXiv:1406.2804 [hep-ph]}}.

\bibitem{Feng:2016emh}
Z.~Feng, G.-M. Huang, and F.~Liu, ``{Anisotropic flow of Pb+Pb $\sqrt{s_{\rm
  NN}}$ = 5.02 TeV from a Multi-Phase Transport Model},''
  \href{http://dx.doi.org/10.1088/1674-1137/41/2/024001}{{\em Chin. Phys. C}
  {\bfseries 41} no.~2, (2017) 024001},
  \href{http://arxiv.org/abs/1606.02416}{{\ttfamily arXiv:1606.02416
  [nucl-ex]}}.

\bibitem{Bilandzic:2010jr}
A.~Bilandzic, R.~Snellings, and S.~Voloshin, ``{Flow analysis with cumulants:
  Direct calculations},''
  \href{http://dx.doi.org/10.1103/PhysRevC.83.044913}{{\em Phys. Rev. C}
  {\bfseries 83} (2011) 044913},
  \href{http://arxiv.org/abs/1010.0233}{{\ttfamily arXiv:1010.0233 [nucl-ex]}}.

\bibitem{Borghini:2000sa}
N.~Borghini, P.~M. Dinh, and J.-Y. Ollitrault, ``{A New method for measuring
  azimuthal distributions in nucleus-nucleus collisions},''
  \href{http://dx.doi.org/10.1103/PhysRevC.63.054906}{{\em Phys. Rev. C}
  {\bfseries 63} (2001) 054906},
  \href{http://arxiv.org/abs/nucl-th/0007063}{{\ttfamily
  arXiv:nucl-th/0007063}}.

\bibitem{Moravcova:2020wnf}
Z.~Moravcova, K.~Gulbrandsen, and Y.~Zhou, ``{Generic algorithm for
  multiparticle cumulants of azimuthal correlations in high energy nucleus
  collisions},'' \href{http://dx.doi.org/10.1103/PhysRevC.103.024913}{{\em
  Phys. Rev. C} {\bfseries 103} no.~2, (2021) 024913},
  \href{http://arxiv.org/abs/2005.07974}{{\ttfamily arXiv:2005.07974
  [nucl-th]}}.

\bibitem{Voloshin:2007pc}
S.~A. Voloshin, A.~M. Poskanzer, A.~Tang, and G.~Wang, ``{Elliptic flow in the
  Gaussian model of eccentricity fluctuations},''
  \href{http://dx.doi.org/10.1016/j.physletb.2007.11.043}{{\em Phys. Lett. B}
  {\bfseries 659} (2008) 537--541},
  \href{http://arxiv.org/abs/0708.0800}{{\ttfamily arXiv:0708.0800 [nucl-th]}}.

\bibitem{Song:2010mg}
H.~Song, S.~A. Bass, U.~Heinz, T.~Hirano, and C.~Shen, ``{200 A GeV Au+Au
  collisions serve a nearly perfect quark-gluon liquid},''
  \href{http://dx.doi.org/10.1103/PhysRevLett.106.192301}{{\em Phys. Rev.
  Lett.} {\bfseries 106} (2011) 192301},
  \href{http://arxiv.org/abs/1011.2783}{{\ttfamily arXiv:1011.2783 [nucl-th]}}.
  [Erratum: Phys.Rev.Lett. 109, 139904 (2012)].

\bibitem{Bhalerao:2014xra}
R.~S. Bhalerao, J.-Y. Ollitrault, and S.~Pal, ``{Characterizing flow
  fluctuations with moments},''
  \href{http://dx.doi.org/10.1016/j.physletb.2015.01.019}{{\em Phys. Lett. B}
  {\bfseries 742} (2015) 94--98},
  \href{http://arxiv.org/abs/1411.5160}{{\ttfamily arXiv:1411.5160 [nucl-th]}}.

\bibitem{Bhalerao:2013ina}
R.~S. Bhalerao, J.-Y. Ollitrault, and S.~Pal, ``{Event-plane correlators},''
  \href{http://dx.doi.org/10.1103/PhysRevC.88.024909}{{\em Phys. Rev. C}
  {\bfseries 88} (2013) 024909},
  \href{http://arxiv.org/abs/1307.0980}{{\ttfamily arXiv:1307.0980 [nucl-th]}}.

\bibitem{Yan:2015jma}
L.~Yan and J.-Y. Ollitrault, ``{$\nu_4, \nu_5, \nu_6, \nu_7$: nonlinear
  hydrodynamic response versus LHC data},''
  \href{http://dx.doi.org/10.1016/j.physletb.2015.03.040}{{\em Phys. Lett. B}
  {\bfseries 744} (2015) 82--87},
  \href{http://arxiv.org/abs/1502.02502}{{\ttfamily arXiv:1502.02502
  [nucl-th]}}.

\bibitem{ALICE:2017fcd}
{\bfseries ALICE} Collaboration, S.~Acharya {\em et~al.}, ``{Linear and
  non-linear flow modes in Pb-Pb collisions at $\sqrt{s_{\rm NN}} =$ 2.76
  TeV},'' \href{http://dx.doi.org/10.1016/j.physletb.2017.07.060}{{\em Phys.
  Lett. B} {\bfseries 773} (2017) 68--80},
  \href{http://arxiv.org/abs/1705.04377}{{\ttfamily arXiv:1705.04377
  [nucl-ex]}}.

\bibitem{Zhou:2015eya}
Y.~Zhou, K.~Xiao, Z.~Feng, F.~Liu, and R.~Snellings, ``{Anisotropic
  distributions in a multiphase transport model},''
  \href{http://dx.doi.org/10.1103/PhysRevC.93.034909}{{\em Phys. Rev. C}
  {\bfseries 93} no.~3, (2016) 034909},
  \href{http://arxiv.org/abs/1508.03306}{{\ttfamily arXiv:1508.03306
  [nucl-ex]}}.

\bibitem{Huo:2017nms}
P.~Huo, K.~Gajdo\v{s}ov\'a, J.~Jia, and Y.~Zhou, ``{Importance of non-flow in
  mixed-harmonic multi-particle correlations in small collision systems},''
  \href{http://dx.doi.org/10.1016/j.physletb.2017.12.035}{{\em Phys. Lett. B}
  {\bfseries 777} (2018) 201--206},
  \href{http://arxiv.org/abs/1710.07567}{{\ttfamily arXiv:1710.07567
  [nucl-ex]}}.

\bibitem{Niemi:2015qia}
H.~Niemi, K.~J. Eskola, and R.~Paatelainen, ``{Event-by-event fluctuations in a
  perturbative QCD + saturation + hydrodynamics model: Determining QCD matter
  shear viscosity in ultrarelativistic heavy-ion collisions},''
  \href{http://dx.doi.org/10.1103/PhysRevC.93.024907}{{\em Phys. Rev. C}
  {\bfseries 93} no.~2, (2016) 024907},
  \href{http://arxiv.org/abs/1505.02677}{{\ttfamily arXiv:1505.02677
  [hep-ph]}}.

\bibitem{Schenke:2020uqq}
B.~Schenke, C.~Shen, and D.~Teaney, ``{Transverse momentum fluctuations and
  their correlation with elliptic flow in nuclear collision},''
  \href{http://dx.doi.org/10.1103/PhysRevC.102.034905}{{\em Phys. Rev. C}
  {\bfseries 102} no.~3, (2020) 034905},
  \href{http://arxiv.org/abs/2004.00690}{{\ttfamily arXiv:2004.00690
  [nucl-th]}}.

\bibitem{ALICE:2018rtz}
{\bfseries ALICE} Collaboration, S.~Acharya {\em et~al.}, ``{Energy dependence
  and fluctuations of anisotropic flow in Pb-Pb collisions at $
  \sqrt{s_{\mathrm{NN}}}=5.02 $ and 2.76 TeV},''
  \href{http://dx.doi.org/10.1007/JHEP07(2018)103}{{\em JHEP} {\bfseries 07}
  (2018) 103}, \href{http://arxiv.org/abs/1804.02944}{{\ttfamily
  arXiv:1804.02944 [nucl-ex]}}.

\bibitem{ALICE:2020sup}
{\bfseries ALICE} Collaboration, S.~Acharya {\em et~al.}, ``{Higher harmonic
  non-linear flow modes of charged hadrons in Pb-Pb collisions at
  $\sqrt{s_{\rm{NN}}}$ = 5.02 TeV},''
  \href{http://dx.doi.org/10.1007/JHEP05(2020)085}{{\em JHEP} {\bfseries 05}
  (2020) 085}, \href{http://arxiv.org/abs/2002.00633}{{\ttfamily
  arXiv:2002.00633 [nucl-ex]}}.

\bibitem{Zhao:2022uhl}
S.~Zhao, H.-j. Xu, Y.-X. Liu, and H.~Song, ``{Probing the nuclear deformation
  with three-particle asymmetric cumulant in RHIC isobar runs},''
  \href{http://dx.doi.org/10.1016/j.physletb.2023.137838}{{\em Phys. Lett. B}
  {\bfseries 839} (2023) 137838},
  \href{http://arxiv.org/abs/2204.02387}{{\ttfamily arXiv:2204.02387
  [nucl-th]}}.

\bibitem{ALICE:2016kpq}
{\bfseries ALICE} Collaboration, J.~Adam {\em et~al.}, ``{Correlated
  event-by-event fluctuations of flow harmonics in Pb-Pb collisions at
  $\sqrt{s_{_{\rm NN}}}=2.76$ TeV},''
  \href{http://dx.doi.org/10.1103/PhysRevLett.117.182301}{{\em Phys. Rev.
  Lett.} {\bfseries 117} (2016) 182301},
  \href{http://arxiv.org/abs/1604.07663}{{\ttfamily arXiv:1604.07663
  [nucl-ex]}}.

\bibitem{ALICE:2021adw}
{\bfseries ALICE} Collaboration, S.~Acharya {\em et~al.}, ``{Measurements of
  mixed harmonic cumulants in Pb\textendash{}Pb collisions at $\sqrt {s_{NN}}$
  = 5.02 TeV},'' \href{http://dx.doi.org/10.1016/j.physletb.2021.136354}{{\em
  Phys. Lett. B} {\bfseries 818} (2021) 136354},
  \href{http://arxiv.org/abs/2102.12180}{{\ttfamily arXiv:2102.12180
  [nucl-ex]}}.

\bibitem{ALICE:2017kwu}
{\bfseries ALICE} Collaboration, S.~Acharya {\em et~al.}, ``{Systematic studies
  of correlations between different order flow harmonics in Pb-Pb collisions at
  $\sqrt{s_{\rm NN}}$ = 2.76 TeV},''
  \href{http://dx.doi.org/10.1103/PhysRevC.97.024906}{{\em Phys. Rev. C}
  {\bfseries 97} no.~2, (2018) 024906},
  \href{http://arxiv.org/abs/1709.01127}{{\ttfamily arXiv:1709.01127
  [nucl-ex]}}.

\end{thebibliography}\endgroup
% \begin{thebibliography}{}
% % and use \bibitem to create references.
% \bibitem{RefJ}
% % Format for Journal Reference
% Author, Journal \textbf{Volume}, (year) page numbers
% % Format for books
% \bibitem{RefB}
% Author, \textit{Book title} (Publisher, place year) page numbers
% % etc
% \end{thebibliography}

\end{document}